\documentclass[10pt,a4paper,twoside]{IEEEtran}

\usepackage[english]{babel}
\usepackage{amsmath}
\usepackage{subfig}
\usepackage{amssymb}
\usepackage{algorithm}
\usepackage{algorithmic}
\usepackage{booktabs}
\usepackage{float} 	
\usepackage{color,soul}
\soulregister\cite7
\soulregister\ref7
\usepackage[dvipsnames]{xcolor}
\usepackage{amsthm}
\usepackage{xcolor}
\usepackage{mathtools, cases}
\usepackage[draft]{hyperref}
\usepackage{lineno}
\usepackage{tabularx}
\usepackage{xparse}
\usepackage[OT1]{fontenc} 

\newcommand{\quotes}[1]{``#1''}
\newcommand{\vect}[1]{\boldsymbol{#1}}
\newcommand{\mat}[1]{\boldsymbol{#1}}
 
\newcommand{\ie}{{i.e.,} }
\newcommand{\eg}{{e.g.,} }

\NewDocumentCommand{\codeword}{v}{\texttt{\textcolor{blue}{#1}}}

\title{Learn to Schedule (LEASCH): A Deep reinforcement learning approach for radio resource scheduling in the 5G MAC layer.}
\author{\IEEEauthorblockN{F. AL-Tam\IEEEauthorrefmark{1}, N. Correia\IEEEauthorrefmark{1}, J. Rodriguez \IEEEauthorrefmark{2}}
\IEEEauthorblockA{\\\IEEEauthorrefmark{1}Centro de Electr\'{o}nica, Optoelectronica e Telecomunica\c{c}\~{o}es (CEOT) - Universidade do Algarve, \\
\IEEEauthorrefmark{2}Instituto de Telecomunica\c{c}\~{o}es, Universidade de Aveiro \\
Email: \IEEEauthorrefmark{1}\{ftam,ncorreia\}@ualg.pt,
\IEEEauthorrefmark{2}jonathan@av.it.pt}}

\begin{document}
\maketitle
\begin{abstract}
	Network management tools are usually inherited from one generation to another. This was successful since these tools have been kept in check and updated regularly to fit new networking goals and service requirements. Unfortunately, new networking services will render this approach obsolete and handcrafting new tools or upgrading the current ones may lead to complicated systems that will be extremely difficult to maintain and improve. Fortunately, recent advances in AI have provided new promising tools that can help solving many network management problems. Following this interesting trend, the current article presents LEASCH, a deep reinforcement learning model able to solve the radio resource scheduling problem in the MAC layer of 5G networks. LEASCH is developed and trained in a sand-box and then deployed in a 5G network. The experimental results validate the effectiveness of LEASCH compared to conventional baseline methods in many key performance indicators. 
\end{abstract}

\section{Introduction}

  The rapid evolution of networking applications will continue to bring new challenges to communication technologies. In the fourth-generation (4G), also known as long term evolution (LTE), throughput and delay were the main foci. In 5G and beyond, services have reached completely new levels. This new era of communication is featured by new killer applications that will benefit from emergent technologies like Internet of things (IoT) and next generation media such as virtual reality (VR) and augmented reality (AR), to name a few.

  Unlike LTE, 5G is a use-case driven technology. In addition, 5G is not only machine-centric but also user-centric, where the user notion has evolved to cover a wider range of entities other than a traditional human-on-handset notion. Small devices that use 5G infrastructure are basically users \cite{ISI:000403536100010}.

  The main use cases supported by 5G, for now, are enhanced mobile broadband (eMBB), ultra-reliable and low latency communications (URLLC) and massive machine-type communications (mMTC). eMMB supports high capacity and high mobility (up to 500 km/h) radio access with 4 ms user plane latency. URLCC provides urgent and reliable data exchange with sub 1 ms user plane latency. The new radio (NR) of 5G will also support massive number of small packet transmissions for mMTC with sub 10 ms latency.

  The main key-enablers to handle the requirements of this new era include flexible numerology, bandwidth parts (BWPs), mini-slotting, optimized frame structure, massive MIMO, inter-networking between high and low bands, and ultra lean transmission \cite{TS_23_501,ISI:000403536100010}. In addition, emergent technologies like software-defined networking (SDN), network function virtualization (NFV), and network slicing will also be key technologies in paving the way for enhancements in 5G and beyond.

  LTE and 5G both rely on the same multi-carrier modulation approach OFDM. Nevertheless, the NR supports multi-numerology structures having different sub-carrier spacings (SCS) and symbol duration. The NR frame structure is more flexible which, on one hand, makes it possible to deliver data for all three main use-cases but, on the other hand, makes it difficult to manage resources efficiently. In addition it is expected that more use cases will emerge and more flexibility is foreseen to be added to the NR frame in the future, making the resource management task even more complicated. For instance, current specifications of the physical layer supports only four BWPs for each user with only one BWP being active at a time. However, UEs in the future will be able to use multiple BWP simultaneously \cite{ISI:000429329400006}.

  The service requirement \cite{ISI:000384887100006}, the significant diversity of the characteristics of the traffic \cite{ISI:000451962400010}, and the user stringent requirements, make 5G a complex system that can not be completely managed by tools inherited from ancestor networks \cite{ISI:000384887100006}. Therefore, industry and academia are looking for novel solutions that can adapt to this rapid growth. One of the main paths is to rely on new AI advancements to solve the network management problems in 5G.

  The current article focuses on a fundamental problem in 5G: the radio resource management (RRM) problem. In general, RRM can be seen as a large problem with many tasks. This article specifically studies the radio resource scheduling (RRS) task in the media access control (MAC) layer. This work shares the same view with many scholars about the necessity of developing AI-based solutions for network management tasks \cite{Feamster:2018:WNR:3232755.3234555}. An important AI tool gaining a lot of attention is the deep reinforcement learning (DRL).  This trend is recently known as learn-rather-than-design (LRTD) approach.

  The main contributions of this work are:
  \begin{itemize}
    \item A numerology-agnostic DRL model;
    \item A clear pipeline for developing/training DRL agents and for their deployment in network simulators;
    \item A comparative analysis in several network settings between the proposed model and the baseline algorithms;
    \item A reward analysis of the model.
  \end{itemize}

  Our approach is novel compared to AI-based approaches which are still scarce. First, this work proposes off-simulator training scheme, which maximizes the flexibility of training the agents and minimizes the training time. Second, our model is tested on an environment different from the one being trained in. From a generalization point of view, we think this should be the case for DRL agents. That is, the training and deployment tasks should be separated. Third, the designed model is new where the state and reward are novel. Finally, our work is tested on a 5G system level simulator that uses all recent components and configurations of a 5G network. Up to our humble knowledge, all these components have not been jointly addressed in any previous work. 

  This article is organized as follows: The RRS problem is described in the reminder of this section. Section \ref{sec:DRL} presents a brief description of the DRL theory. The related work is presented later in Section \ref{sec:related_work}. Sections \ref{sec:LEASCH} and \ref{sec:results} present the proposed approach and the results, respectively. The article is then concluded in Section \ref{sec:conclusions}.

  \subsection{Radio resource scheduling problem }
    The continuous update of physical layers to handle new use cases in communications is the main surge behind the development of flexible MAC layers or components thereof. As new use cases emerge, handcrafted MAC layers become more complicated and prone to error. This is, in fact, the main problem in modern network and resource management \cite{Feamster:2018:WNR:3232755.3234555,ISI:000473097800013}. Human-centered approaches lack flexibility and usually require continuous repairs and updates, which lead to a degradation in the level of service and compromise in the performance \cite{ISI:000465405500021}.

    Improving the ability of communication systems, to effectively share the available scarce spectrum among multiple users, has always been the main research target of academia and big com industry. As more user requirements are added to the system, the need to find better resource sharing approaches becomes inevitable. Therefore, RRS is an essential task in communications. The main objective of RRS task is to dynamically allocate spectrum resources to UEs while achieving an optimal trade-off between some key performance indicators (KPIs), like spectrum efficiency (SE), fairness, delay, and so on \cite{ISI:000318893100007}. Achieving such trade-off is known to be a combinatorial problem. 

\begin{figure}
    \centering
    \includegraphics[width=\columnwidth, clip]{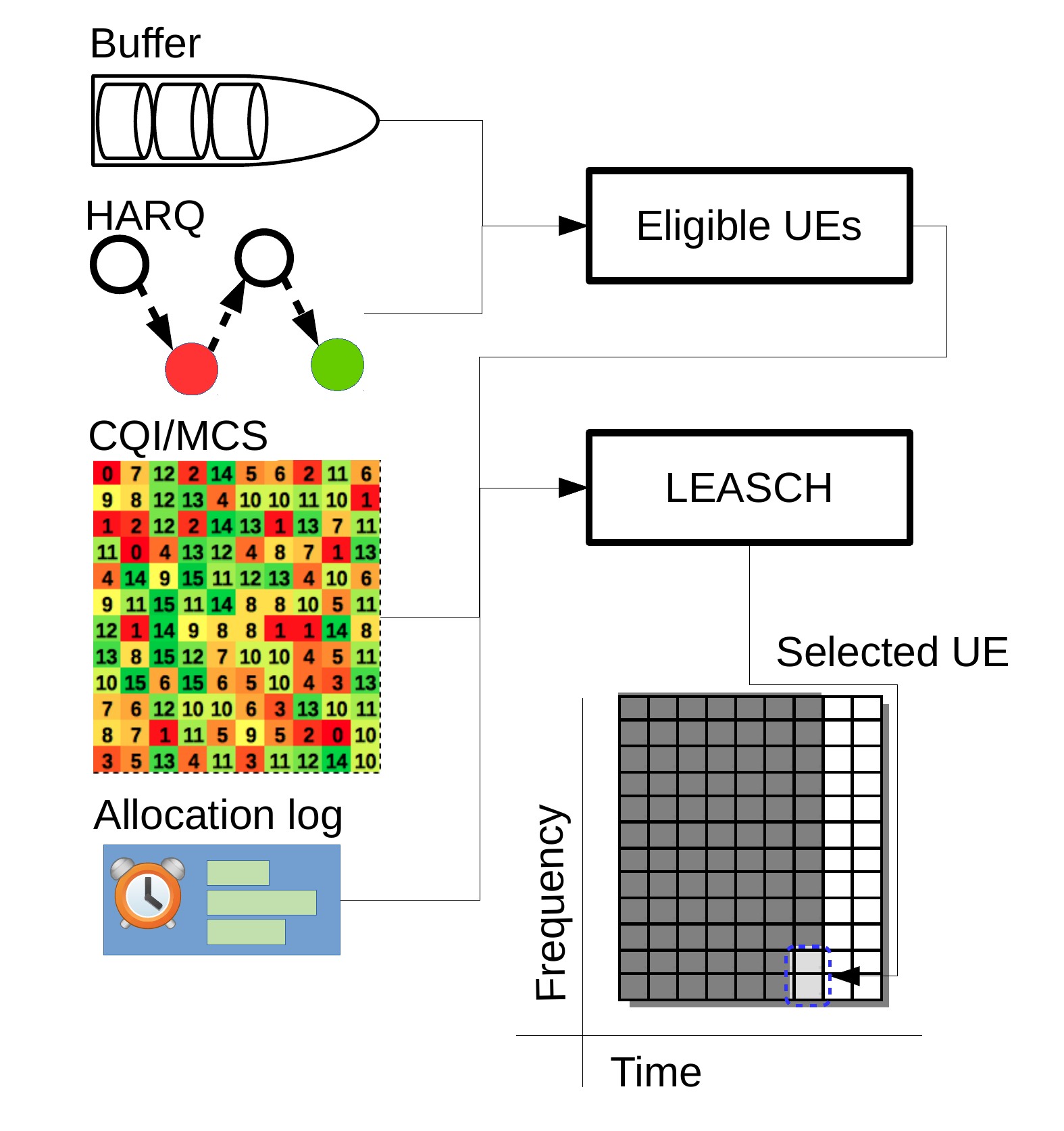}
    \caption{\label{fig:LEASCH}RBG scheduling with LEASCH.}
  \end{figure}

    As described in Figure \ref{fig:LEASCH}, RRS simply works as follows: the scheduler runs at the gNB at every (or $k$th) slot to grant the available resource block groups (RBGs) to the active UEs. Therefore, the problem boils down to filling the resource grid by deciding which UE will win the current RBG in the current slot. However, not all users can be scheduled at the current RBGs. Only those that are eligible (active) will be considered for scheduling and, therefore, allowed to compete for the RBGs under consideration. A UE is eligible if it has data in the buffer and is not retransmitting in the current slot, \ie if it is not associated with a HARQ process in progress.

    In many cases, obtaining an optimal solution for RRS problem is computationally prohibitive due to the size of the state-space and the partial observability of the states \cite{8532121}. Moreover, surged by new requirements, the RRS task will continue to expand, in the future, both horizontally and vertically. Horizontally, regarding the number and diversity of users, and traffic patterns it should support, and vertically by having to consider new (and perhaps contradictory) KPIs. Therefore, RRS can easily become intractable even for small-scale scenarios.

    Current RRS solutions are driven by conventional off-the-shelf designated tools. This includes variants of the proportional fairness (PF), round robin (RR), BestCQI, and so on. This scheme has been successful but it will become obsolete in the future. In fact, this is the case of many current network management tasks \cite{ISI:000484917800217}. A new RRS approach is inevitable due to:
    \begin{itemize}
     \item the rapid increase in network size;
     \item the breadth of control decisions space; 
     \item the new perception from business-makers and end-users of the networking services and applications;
     \item modern networks are delayed return environments; 
     \item lack of sufficient understanding of network underlying by conventional tools, i.e., they are myopic. 
   \end{itemize}



    In this context, we share the same vision with \cite{Feamster:2018:WNR:3232755.3234555} that research communities and industry have been focusing on developing services, protocols, and applications more than developing efficient management techniques. Fortunately, recent technologies in artificial intelligence (AI) offer promising solutions and there is a consensus among many scholars of the need of AI-powered network management \cite{ISI:000483018200007}.

  \section{Deep Reinforcement Learning} \label{sec:DRL}

     RL is a learning scheme for sequential decision problems to maximize a cumulative future reward. An environment of such scheme can be modeled as a Markov decision process (MDP) represented by the tuple $(\mathcal{S}, \mathcal{A}, \mat{P}, r, \gamma)$. Where $\mathcal{S}$ is a compact space of states of the environment. $\mathcal{A}$ is a finite set of possible actions (action space), $\mat{P}$ is a predefined transition probability matrix such that each element $p_{ss^{\prime}}$ determines the probability of transition from state $s$ to $s^{\prime}$. The reward function  $r:\mathcal{S} \times \mathcal{A} \times \mathcal{S} \rightarrow \mathbb{R}$ tells how much reward the agent will get when moving from state $s$ to state $s^{\prime}$ due to taking action $a$. The $\gamma$ is a discount factor used to trade-off the importance between immediate and long-term rewards. 


     In RL, an agent learns a policy $\pi$ by interacting with environment. In each time step $t$, the agent observes a state $s_t$, takes an action (decision) $a_t$, observes a new state $s_{t+1}$ and receives a reward signal $r(s_t, a_t)$. The learning scheme can be episodic or non-episodic and some states are terminal.

    A policy $\pi$ is a behavioral description of the agent and the policy for state $s$, $\pi(s)$, can be defined as a probability distribution over the action space $\mathcal{A}$, such that the policy for the pair $(s, a)$, $\pi(a|s)$, defines the probability assigned to $a$ in state $s$. Therefore, a policy simply tells us which action to take at state $s$.

    The objective of training an agent is to find an optimal policy that will tell the agent which action to take when in a specific state. Therefore, the objective of an agent boils down to maximizing the expected reward for a long run. Starting from state $s_t$, the outcome (return) can be expressed as:
   \begin{equation}
    G_t = \mathbb{E}\left[\sum\limits_{k = 0}^{\infty} \gamma^k r(s_{t+k}, a_{t + k})| s_0 = s_t\right]
   \end{equation}
    For a non-episodic learning scheme, we can see that $\gamma < 1$ is important not only to obtaining a trade-off between immediate and long-term rewards but also for mathematical convenience.

    When an agent arrives at a state it needs to know how good it is to be at state $s$ and following the optimal policy afterwards. A function to measure that is called the value, aka state-value, function $V(s)$:
     \begin{equation}
      V(s) = \mathbb{E}\left[G_t | s_t = s \right]
     \end{equation}

     Similarly, to measure how good it is to be at state $s$ and take action $a$, a quality function $Q$, aka action-value function, can also be derived as:
     \begin{equation}
      Q(s, a) = \mathbb{E}\left[G_t | s_t = s, a_t = a\right]
     \end{equation}

     Once we know $Q$ and $\pi$ we can calculate $V$ using:

     \begin{equation}
      V(s) = \sum_{a \in \mathcal{A}} \pi(a|s) Q(s, a)
     \end{equation}

     Therefore, $V$ and $Q$ can be related as:
     \begin{equation}
          V(s) = \mathbb{E}_{a \sim \pi(a|s)} [Q(s, a)].
     \end{equation}In addition, these two functions can also be related via an advantage function $A$ \cite{10201}:
     \begin{equation}
      A(s, a) = Q(s, a) - V(s),
     \end{equation}
     where $A$ subtracts the value function from the quality function to obtain a relative importance of each action, and tell the agent if choosing an action $a$ is better than the average performance of the policy.

     In fact, we are interested in finding $Q$ since we can easily derive the optimal policy $\pi^*$ from the optimal $Q^*$. $Q(s, a)$ maps each $(s, a)$ pair to a value, \ie it measures how good it is to take action $a$ when  in state $s$ and then following the optimal policy. Using the Bellman expectation function, we can rewrite $Q(s, a)$ as:
     \begin{equation}
      Q(s, a) = r(s, a) + \gamma \sum\limits_{s^\prime \in \mathcal{S}} p_{ss^\prime}(a) V(s^\prime) 
     \end{equation}
      Therefore, following the Bellman optimality equation for $Q^*$ we have:

      \begin{equation}
      Q^*(s, a) = r(s, a) + \gamma \sum\limits_{s^\prime \in \mathcal{S}} p_{ss^\prime}(a) \max\limits_{a^\prime} Q^*(s^\prime, a^\prime)
      \end{equation}

      The optimal policy can be then derived from the optimal values $Q^*(s, a)$ by choosing the maximum action value in each state. This scheme is known as value-based (compared to policy-based) learning since the policy is driven from the value function:

      \begin{equation}
        \pi^*(s) = \arg\max\limits_{a \in \mathcal{A}} Q^*(s, a), \quad \forall s\in\mathcal{S}
      \end{equation}

      However, finding $\pi^*$ is not easy since in may real world applications, the transition probability is not known. One algorithm to solve this Bellman optimality equation is the Q-learning algorithm \cite{q_learning}. This algorithm is off-policy critic-only (compared to on-policy and actor-critic algorithms). In this algorithm, $Q$ is represented as a lookup table, which can be initialized by random guesses and gets updated in each iteration using the Bellman Equation:
    \begin{equation}
      Q(s_t, a_t) = r(s_t, a_t) + \gamma \max\limits_{a_{t+1}} Q(s_{t+1}, a_{t+1})
    \end{equation}
    
    For terminal state this update comes down to:
        \begin{equation}
      Q(s_t, a_t) = r(s_t, a_t)
    \end{equation}
    In order to balance between exploration and expedition, the agent, in Q-learning, adapts an $\epsilon$-greedy algorithm. In $\epsilon$-greedy, the agent selects an action $a$ using $a = \arg\max\limits_{a^\prime \in \mathcal{A}} Q(s, a^\prime)$ with probability $1-\epsilon$, otherwise selects a random action with probability $\epsilon$. This randomness in decision making helps the agent to avoid local minimums. As the agent progresses in learning, it reduces $\epsilon$ via a decaying threshold $\delta_\epsilon$. With this annealing property of $\epsilon$-greedy, in practice, an agent is expected to perform almost randomly in the beginning and matures with time.

    One drawback of the original Q-learning algorithm is scalability. Keeping a tabular for such iterative update is feasible only for small problems. For larger problems, it is infeasible to keep track of each $(s, a)$ pairs. Therefore, in practice it is more feasible to approximate $Q$.

    A common way to approximate $Q$ is to use a deep neural network (DNN). This cross-breeding between deep learning and Q-learning has yielded deep Q networks (DQN), more generally known as deep reinforcement learning (DRL), which is the main breakthrough behind recent advancements in RL that delivered a human-level performance in Atari games \cite{ISI:000350097300045} and even more strategic games \cite{ISI:000496938200057} where the agent learns directly from a sequence of image frames via convolutional neural networks (CNN) and DRL. 

    In DQN, the $Q$ function is approximated by minimizing the squared error between the Bellman equation and the neural network estimation, aka mean-squared Bellman error (MSBE):
    \begin{equation}
      \text{loss} = \left( Q(s_t, a_t; \vect{\theta}) - Q^{\text{target}} \right)^2
      \label{equ:e}
    \end{equation}
    where $Q^{\text{target}}$ is the target $Q$ function, known as the target critic, and $\vect{\theta}$ is the set of DNN parameters.  $Q^{\text{target}}$ is calculated as:
     \begin{equation}
      Q^{\text{target}} = r(s_t, a_t) + \gamma \max\limits_{a_{t+1}} Q(s_{t+1}, a_{t+1}; \vect{\theta})
     \end{equation}
     where $\vect{\theta}$ is the set of DNN's weights and is updated in a stochastic gradient descent (SGD) fashion. For a predefined learning rate $\alpha$ and a mini-batch size $M$,  $\vect{\theta}_t$ is updated using:
    \begin{equation}
      \vect{\theta}_t = \vect{\theta}_t+ \frac{\alpha}{M} \left( Q(s, a; \vect{\theta}_t) - Q^{\text{target}}(\vect{\theta}_t) \right) \nabla_{\vect{\theta}_t} Q(s, a; \vect{\theta}_t) 
      \label{equ_update_theta}
    \end{equation}
    where $\left( Q(s, a; \vect{\theta}_t) - Q^{\text{target}}(\vect{\theta}_t) \right)$ is known as the temporal difference (TD) error.

    In DQN, the state and actions are represented by two separate networks and combined via an \texttt{Add} layer. The output is a single value ($Q$ value) in a way similar to classical regression. However, a more efficient architecture is to have the state as input and let the network output be equal to the length of action space. This way, each output represents the likelihood of an action given the state. As in classical Q-learning, the action with maximum likelihood will be selected. 

    In order to stabilize the results, and to break any dependency between sequential states, DQN uses two tricks. First, two identical neural networks are used one for on-line learning and another to calculate the target $Q^{\text{target}}$. The target network is updated periodically, from the on-line network, every $T$ steps. Therefore, the target is calculated from a more mature network, thus increasing the learning stability:

     \begin{equation}
      Q^{\text{target}} = r(s_t, a_t) + \gamma \max\limits_{a_{t+1}} Q(s_{t+1}, a_{t+1}; \vect{\hat{\theta}})
     \end{equation}
     where $\vect{\hat{\theta}}$ is a delayed version of $\vect{\theta}$

     Instead of copying the weights from the on-line to the target network at every $T$ steps, it turns out that a smoothing (\ie progressive) update approach can noticeably increase the learning stability: 
    \begin{equation}
      \hat{\vect{\theta}} = \beta \vect{\theta} + (1 - \beta) \hat{\vect{\theta}}
      \label{equ:smoothing_update}
    \end{equation}
    where $\beta$ is a small real-valued smoothing parameter.

    The second trick is to use an experience replay memory $\mathcal{R}$, usually implemented as a cyclic queue. This memory is updated in every learning step, by appending the tuple  $(s, a, s_{t+1}, r_{t+1})$ to the end of the queue. Therefore, when training $Q$, random mini-batches are sampled from $\mathcal{R}$ and fed to the $Q$ on-line network. $\mathcal{R}$ reduces the dependency between consequence input and improve the data efficiency via re-utilizing the experience samples. 

    Q-learning and its variant DQN tend to be overoptimistic due to the noise in the $Q$ estimates, and the use of the max operator in selecting the action and calculating the value of the action. A solution is the Double DQN (DDQN) \cite{NIPS2010_3964,ISI:000485474202019} model, which learns from two different samples. One is used to select the action and another one is used to calculate the action value. Therefore, in DDQN, the critic target $Q^{\text{target}}$ is calculated as:

    \begin{equation}
      Q^{\text{target}} = r_{t+1}(s, a) + \gamma  Q(s_{t+1}, \arg\max\limits_a Q(s_{t+1}, a, \vect{\theta}); \vect{\hat{\theta}})
      \label{equ:DDQN_update}
    \end{equation}

    In expression \eqref{equ:DDQN_update}, the selection of the action is made from the on-line network, \ie $\arg\max\limits_a Q(s_{t+1}, a, \vect{\theta})$, and the evaluation and update is made from the target critic network $Q(s_{t+1}, \arg\max\limits_a Q(s_{t+1}, a, \vect{\theta}); \vect{\hat{\theta}})$.

 \subsection{Why DRL is suitable for RRM problem?}
  
    The notion of \emph{self-driving} networks \cite{Feamster:2018:WNR:3232755.3234555} is gaining more and more attention nowadays. The core vision of self-driving network engineering is to  learn rather than to design the network management solutions \cite{Joseph:2019:TSR:3301293.3302374}. Such vision has radically changed some fields like computer vision via deep learning \cite{ISI:000402555400026}, by learning features rather than hand-crafting them. However, we did not witness such major progress in network management. The reason is that supervised learning is not suitable for some control and decision problems, since collecting and labeling networking data is not trivial, expensive, and network states are non-stationary \cite{Benson:2010:NTC:1879141.1879175,ISI:000485488900094}. DRL can be quite suitable for such problems due to the following reasons:

  \begin{itemize}
    \item All information about RRM can be centralized in the gNB thus creating a network wide view (although not fully) of the network. In addition, new paradigms like knowledge defined networking (KDN) can be used \cite{ISI:000413345900005};
    \item DRL agents can continue learning and improving while the network operates. They can interact with other conventional components in the system and learn from them if necessary \cite{wang2019deep};
    \item Network dynamics are difficult to anticipate and exact mathematical models are not scalable. For 5G network management, it is extremely difficult to model the network state and traffic due to the diversity of the applications and traffic it supports \cite{ISI:000451962400010}. Therefore, DRL model-free models can be the choice in this case.
    \item After the new breakthroughs, DRL becomes an extremely hot research topic \cite{ISI:000485488903036}. In networking, the popularity of DRL is increasing and some famous network simulators recently have been extended to support general DRL environments like gym \cite{Gawlowicz:2019}.
  \end{itemize}


  \section{Related work} \label{sec:related_work}

    DRL solutions for RRS are scarce in the literature but they can be divided according to the nature of the action space into two main categories: Coarse (high-level) and fine-grained (low-level) decisions. In the former, the DRL agent acts as a method/algorithm selector \cite{ISI:000428054305037,ISI:000484917800217} or protocol designer \cite{ISI:000465405500021,ISI:000473097800013}. For instance, for a given network state, the DRL agent selects which conventional algorithm is suitable to perform scheduling. In the latter, DRL decisions are hard-wired in the networking fabric. The DRL agent makes fine-grained decisions like filling the resource grid \cite{wang2019deep}, air-time sharing between users and decide which user has rights to access the channel \cite{8532121,8359094}, or select which coding scheme is suitable \cite{ISI:000471120800028}. In addition, the fine-grained methods can be classified into distributed \cite{8532121,ISI:000468234400008} and centralized \cite{8359094,wang2019deep}. In the distributed approaches, each UE acts as a DRL agent. This way the network is composed of multi-agents in a way similar to those in game theory. Such approach is scalable but sharing the network state among multiple entities makes it difficult to guarantee convergence. On the other hand, centralized approaches can benefit from a better computational power and better network state understanding.

    Both coarse and fine-grained approaches have pros and cons. The coarse level scheme is more scalable, since the agent acts in almost a constant action space. On the other hand, such approach falls short in obtaining deep control of the network. Conventional algorithms are still the main working horses. In fine-grained approaches, the DRL agent deals with the finest decisions. Therefore, it can obtain a deep control of the network. However, these approaches require more sophisticated designs to be adaptive to networking dynamics. Our work belongs to the fine-grained centralized approaches.

    \subsection{Coarse approaches}

    An algorithm selector approach can be found in \cite{ISI:000428054305037}. At each slot, an actor-critic agent chooses a scheduling algorithm, among a set of available PF-variants algorithms, to maximize some QoS objectives. The state is the number of active users, the arrival rate, the CQIs, and the performance indicator with respect to the user requirements. The reward function measures the impact of choosing a rule on the QoS satisfaction of the users. A similar approach can be found in \cite{ISI:000484917800217} for 5G network but using a variant of actor-critic DRLs known as deep deterministic policy-gradient (DDPG) algorithm and with larger action space that controls more parameters. However, this approach is not numerology-agnostic. In \cite{ISI:000484917800217}, for instance, a distinct DRL design is required for each network setting.

    In \cite{ISI:000465405500021}, AlphaMac is proposed which is a MAC designer framework that selects the basic building blocks to create a MAC protocol using a constructive design scheme. A building block is included in the protocol if its corresponding element in the state is 1, zero otherwise. As action, the agent chooses the next state that will increase the reward (which is the average throughput of the channel). Each selection by the agent is then simulated in an event-driven simulator that mimics the MAC protocol but with flexibility to allow adding and removing individual blocks of the protocol.

    Physical layer self-driving radio is proposed in \cite{Joseph:2019:TSR:3301293.3302374}. The user specifies the control knobs, and other requirements, and the system learns an algorithm that fits a predefined objective (reward) function. The action space is the control knobs and their possible settings. The system then holds a set of DNN and applies the appropriate one to the input scenario. In fact this work can be regarded as hybrid since it combines both coarse and fine-grained approaches in a hierarchal design.

    \subsection{Fine-grained approaches}
    A general resource management problem is handled in \cite{ISI:000390598000008} by a policy gradient DRL agent. The objective is to schedule a set of jobs at a resource cluster at a given time step. This work demonstrated the suitability of DRL agents in one hand but, on the other hand, it can not be applied directly to 5G RRS problems. 

    A RRS agent for LTE networks can be found in \cite{wang2019deep}. A single RBG is considered and the authors have shown that DRL agent, trained by the DDPG algorithm, can achieve near PF results when it uses PF algorithm as an expert (guide) to learn from. This approach can ensure great stability since the agent learns from a well-established algorithm, but it diminishes the ability of agents to discover their own policies.


    In \cite{ISI:000485488900094} a high volume flexible time (HVFT) traffic driven by IoT is scheduled on radio network via a variant of DDPG algorithm, where the scheduler determines the fraction of IoT traffic on top of conventional traffic. To empower the agent with time notion, a temporal features extractor is used, and these features are then fed to the agent. The reward function is a linear combination of several KPIs, like IoT traffic served, traffic loss due to the introduction of IoT traffic and the amount of served bytes below as the system-wide desired limit.

    In \cite{8359094} a policy gradient DRL is proposed to manage the resource access between LTE-LAA small base stations (SBS) and Wi-Fi access points. The goal is to determine the channel selection, carrier aggregation, and fractional spectrum access for SBS while considering airtime fairness between SBS and WI-FI APs. The state includes all network nodes states, and the reward is the total throughput over the selected channels. The scheduling problem is modeled as a non-cooperative Homo Equalis game model where, in this model, the achievement of a player is calculated by its performance while maintaining a certain fair equilibrium regarding other players. To solve this model and establish a mixed strategy, a deep learning approach is developed, where LSTM and MLP networks are used to encode the input data (from IBM Watson Wi-Fi data set) and the objective function of the model is solved via a REINFORCE-like algorithm. The work has shown improvement in the throughput compared to reactive RL, when increasing the time horizon parameter. In addition, when compared to classical scheduling approaches like PF, the work shows enhancement in served network traffic but at the same time the average airtime allocation for Wi-Fi APs has degraded as the time horizon parameter increases. One disadvantage of this work is that it uses a heavy-weight architecture.

    In \cite{8532121} a lightweight multi-user deep RL approach is used to address spectrum access problem where a recurrent Q network (RQN) \cite{hausknecht2015deep} with dueling \cite{10201} is used. At each time slot a user can only select a single channel to transmit, and if the transmission is successful then an ACK signal (observation) is sent back to the user, otherwise a collision has happened. When modeling this problem in an RL framework, the length of the action space of a user is a $|\mathcal{C}| + 1$ binary vector (one-hot), and $\mathcal{C}$ is the set of channels, indicating which channel was selected by the user. The first element of this vector is 1 if the user has decided to wait. The state is the history of actions and the observations made by a user $u$ until time $t$. The reward is the achievable data rate.
    The training phase of this work is centralized, while the deployment phase is distributed, and the model weights are updated in each UE only when required, \eg after substantial change in the UE behavior.

    In \cite{8761566}, the duty cycle multiple access mechanism is used to divide the time frame between LTE and Wi-Fi users. A DLR approach is then used to find the splitting point based on the feedback averaged from the channel status for several previous frames. Information like idle slots, number of successful transmissions, action, reward are used to represent the state of the agent. The action is a splitting point in the time frame (\ie an integer), and the reward is the transmission time given to the LTE users while not violating the Wi-Fi users minimum data rate limit.

    In \cite{ISI:000468234400008} a DQN model is developed to learn how to grant access between an DRL agent and different infrastructures. The agent learns by interacting with users that use other protocols, like TDMA and ALOHA, and learns to send its data in the slots where the other users are idle.



  \section{The proposed DRL scheduler (LEASCH)} \label{sec:LEASCH}
    A general sketch of the proposed scheduler is shown in Figure \ref{fig:LEASCH}. Our work has two phases. Training and testing. In the training, the scheduling task is transformed into an episodic DRL learning problem and LEASCH is trained until it converges. In the testing phase, a 5G system level simulator is used to deploy LEASCH. These two phases are described deeply in the following subsections. Each component of LEASCH is described from a DRL perspective first, and then the training and deployment algorithms are presented:
    
    \subsection{LEASCH's design}
    \subsubsection{State} Let us recall the objective of our agent as a scheduler. At a given RBG, it has to select an active (eligible) UE from a set of candidate UEs and assign that RBG to the selected UE. Our objective is to jointly optimize the throughput and fairness. Therefore, we can divide our state into three parts: eligibility, data rate, and fairness. We derive each part separately and then combine them in a single input vector representing the state.


     \paragraph{Eligibility} At each RBG there is only a subset of UEs, $\hat{\mathcal{U}}\subseteq \mathcal{U}$, eligible for the current RBG. A user is eligible for scheduling at a given RBG if the UE has data in the buffer and is not associated with a HARQ process. However, instead of feeding the buffer and the HARQ status of each UE to the LEASCH, and ask the agent to learn \quotes{eligibility}, we simplify the task for the agent by calculating a binary vector $\vect{g}$ to act as an eligibility indicator:
    \begin{equation}
    g_u = \left\{ \begin{array}{ll}
                  1, & \text{ if } u \text{ is eligible} \\
                  0, & \text{Otherwise}  
                  \end{array}\quad ,\forall u \in \mathcal{U} \right.
                  \label{equ:g}
    \end{equation}

    As we will see, $\vect{g}$ will help us designing a tangible reward function that allows the agent to effectively learn how to avoid scheduling inactive UEs.

    \paragraph{Data rate}
    One way to represent this piece of information in the agent state is to use the data/bit rate directly. However, we use the valid entries of modulation and coding schemes (MCSs) in Table 5.1.3.1-2 in the 5G physical layer specification TS 38.214 \cite{TS_38_21} to model this information. We denote to this information vector by $\vect{d}$.


    \paragraph{Fairness}
    We keep track each time a UE is admitted to an RBG. To that end, a vector with all-zero elements $\vect{f} = \vect{0}$ is created in the beginning of each episode, and each time an RBG is scheduled $\vect{f}$ is updated:
    \begin{equation}
      f_u = \left\{ \begin{array}{ll}
                  \max(f_u - 1, 0), & \text{if } u \text{ is selected} \\
                  f_u + 1, & \text{if } u  \text{'s buffer is not empty}  
                  \end{array},\forall u \in \mathcal{U} \right.
                  \label{equ:f}
    \end{equation}
    Therefore, $\vect{f}$ represents the \emph{allocation-log} of the resources. In the best case scenario, all entries of $\vect{f}$ are the same, meaning that all UEs are admitted to the resources with the same probability. In addition, $\vect{f}$ also represents the delay, because if a UE did not access the resources for too long, its corresponding value in $\vect{f}$ will be large. 

    Combining these three vectors $\vect{g}, \vect{d}$ and $\vect{f}$ yields the state. The size of the state can be further reduced by joining $\vect{g}$ and $\vect{d}$ via the Hadamard product:
    \[\hat{\vect{d}} = \vect{d} \circ \vect{g}\]
    making the final state vector defined by:
    \begin{equation}
      \vect{s} = \left[\hat{\vect{d}} \quad \vect{f}\right]^\top 
      \label{equ:state}
    \end{equation}

    This way our state represents all pieces of information in a compact but descriptive manner. For a better learning stability we normalize $\hat{\vect{d}}$ and $\vect{f}$ to the range $[0, 1]$.
  \subsubsection{Action}
    The action is to select one of the UEs in the system. It is encoded in hot-one encoding.

  \subsubsection{Reward}
    Reward engineering is a key problem in RL in general. In general, the reward is treated similarly to an objective function to be maximized. However, we believe that it should be engineered as a signal such that each state-action pair represents a meaningful reward. 

    From our state design the goal is to encourage the agent to transmit at the RBGs with the highest MCS, \ie highest bit-per-symbol, to increase the throughput in the system. At the same time, we would like the agent not to compromise the resource sharing between the users. Therefore, the adopted reward is given by:

    \begin{equation}
      r(s, u; K) = \left\{
                  \begin{array}{ll}
                   - K, & \text{if } u \text{ is none-eligible} \\
                   \hat{d}_u \times \frac{\min\limits_{u} f_u}{\max\limits_{u} f_u}, & \text{otherwise}
                  \end{array}
                \right.
                \label{equ:reward}
    \end{equation}
    where $K$ is a threshold to represent the negative penalization signal for scheduling an inactive UE, and $\vect{f}$ is updated using \eqref{equ:f}. We can easily see that, our reward is a variant of a discounted bestCQI function, where the data rate is discounted by the resource sharing fairness.
    \subsection{LEASCH training and deployment}
    LEASCH is trained for a sequence of episodes. The training procedure of one episode is described in Algorithm \ref{alg:train_leasch}.  In the beginning of each episode, a random state is created. Then the agent is trained for a set of $\ell_{\text{episode}}$ steps. In each step the agent trains its on-line $Q$ neural network, and transfers the learned parameters to the target critic neural network each $T$ steps. After an episode has finished, the experience replay memory $\mathcal{R}$ and the learned weights are transfered to the next episode, and so on. The state is reseted in the beginning of each episode. 

    Once the training phase has finished, LEASCH is deployed in a 5G simulator to test it. The deployment algorithm is shown in Algorithm \ref{alg:deploy_leasch}. In this algorithm, the agent is plugged in like any other conventional scheduling algorithm. Each time an RBG is ready for scheduling, it is admitted to LEASCH which first calculates the set of eligible UEs, $\hat{\mathcal{U}}$, and create a state $\vect{s}$. Next, it decides which UE wins the RBG by performing a forward step on its neural network with weights $\vect{\theta}$ and chooses the action with the highest probability. If the selected UE, $u$, belongs to $\hat{\mathcal{U}}$ then LEASCH assigns the current RBG to $u$. According to LEASCH's decision, the simulator allocates the resources and records statistics.

\begin{algorithm}[]
\begin{algorithmic}[1]
\STATE // \emph{input: $\ell_\text{episode}$, $K$, $M$, $T$ $\epsilon$, $\delta_\epsilon$, $\min_\epsilon$, $\vect{\theta}$, $\hat{\vect{\theta}}$, $\mathcal{R}$.}
\STATE // \emph{output: updated \{$\vect{\theta}, \hat{\vect{\theta}}$, $\mathcal{R}$\}.}
	  
	  \STATE initialize $\vect{s}$ randomly according to the ranges of $\hat{\vect{d}}$ and $\vect{f}$ 
    \FOR {$i = 1:\ell_\text{episode}$}  
      \STATE forward $\vect{s}$ to the on-line Q neural network and get the selected UE, $u$, via $\epsilon$-greedy as: \\
      			$u = \arg\max\limits_{a \in \mathcal{A}} Q(\vect{s}, a; \vect{\theta})$

      \STATE anneal $\epsilon$ as: $\max\{ \epsilon - \delta_{\epsilon}, \min_{\epsilon}\}$
      \STATE calculate the reward $r(\vect{s}, u; K)$ using \eqref{equ:reward}.
      \STATE calculate new state $s^\prime$ using the equations \eqref{equ:g} to \eqref{equ:state}
      \STATE add the tuple $(\vect{s}, u, r, \vect{s^\prime})$ to the experience replay $\mathcal{R}$
      \STATE sample $M$ mini-batches from $\mathcal{R}$ and train the on-line Q neural network with $\vect{\theta}$ using \eqref{equ_update_theta} and \eqref{equ:DDQN_update}
      \STATE update the target critic Q neural network (with $\hat{\vect{\theta}}$) using $\vect{\theta}$ every $T$ steps via smoothing \eqref{equ:smoothing_update}. 
      \STATE $\vect{s}\gets \vect{s^\prime}$ 
    \ENDFOR

\RETURN \{$\vect{\theta}, \hat{\vect{\theta}}$, $\mathcal{R}$\}
\end{algorithmic}
\caption{- \textbf{Training phase of LEASCH}.}\label{alg:train_leasch}
\end{algorithm}

\begin{algorithm}[]
\begin{algorithmic}[1]
\STATE // \emph{input: trained LEASCH.}
\FOR {\textbf{each} time slot} 
    \FOR {\textbf{each} RBG}  
      \STATE calculate the set of eligible UEs $\hat{\mathcal{U}}$  
      \IF {$\hat{\mathcal{U}} \neq \emptyset$} 
      	\STATE calculate state $\vect{s}$
      	\STATE forward $\vect{s}$ to LEASCH 
      	\STATE calculate the action $u$ as: \\  $u = \arg\max\limits_{a \in \mathcal{A}} Q(s, a; \vect{\theta})$  
      	\IF {$u \in \hat{\mathcal{U}}$}
      		\STATE schedule $u$ for the current RBG 
      	\ENDIF
      \ENDIF
      \STATE collect statistics from the simulator
    \ENDFOR
\ENDFOR

\end{algorithmic}
\caption{- \textbf{Deployment phase of LEASCH in 5G.}}\label{alg:deploy_leasch}
\end{algorithm}

\section{Results} \label{sec:results}
  In order to evaluate the proposed scheduler, a comparison with two baseline algorithms, proportional fairness (PF) and round robin (RR). These are widely used algorithms in literature and in practice. The main objective here is to assess LEASCH using different settings in order to: \emph{i}) show its ability to solve the RRS problem; \emph{ii}) try to understand which policy it was able to learn; and \emph{iii}) to analyze the quality of its design. The collected results were analyzed from different perspectives in order to accomplish these goals.

 \subsection{Experimental setup}
 	The parameters adopted for LEASCH and 5G simulator are depicted in Tables \ref{tbl:leasch_params} and \ref{tbl:simulator_params}, respectively. As for LEASCH's architecture, its $Q$ neural networks are DNNs with two fully connected hidden layers of 128 neurons each, and \texttt{relu} activation functions. The input layer size is $2 \times |\mathcal{U}|$ while the output layer is a layer of size $|\mathcal{U}|$. 

 	All methods and algorithms presented/discussed here are implemented in Matlab 2019b in a PC running Linux with i7 2.6GHz, 32GB RAM, and GPU Nvidia RTX 2080Ti with 11 GB. 

\begin{figure}
 	\centering
    \includegraphics[width=0.9\columnwidth, trim=0cm 0cm 0cm 0.45cm, clip]{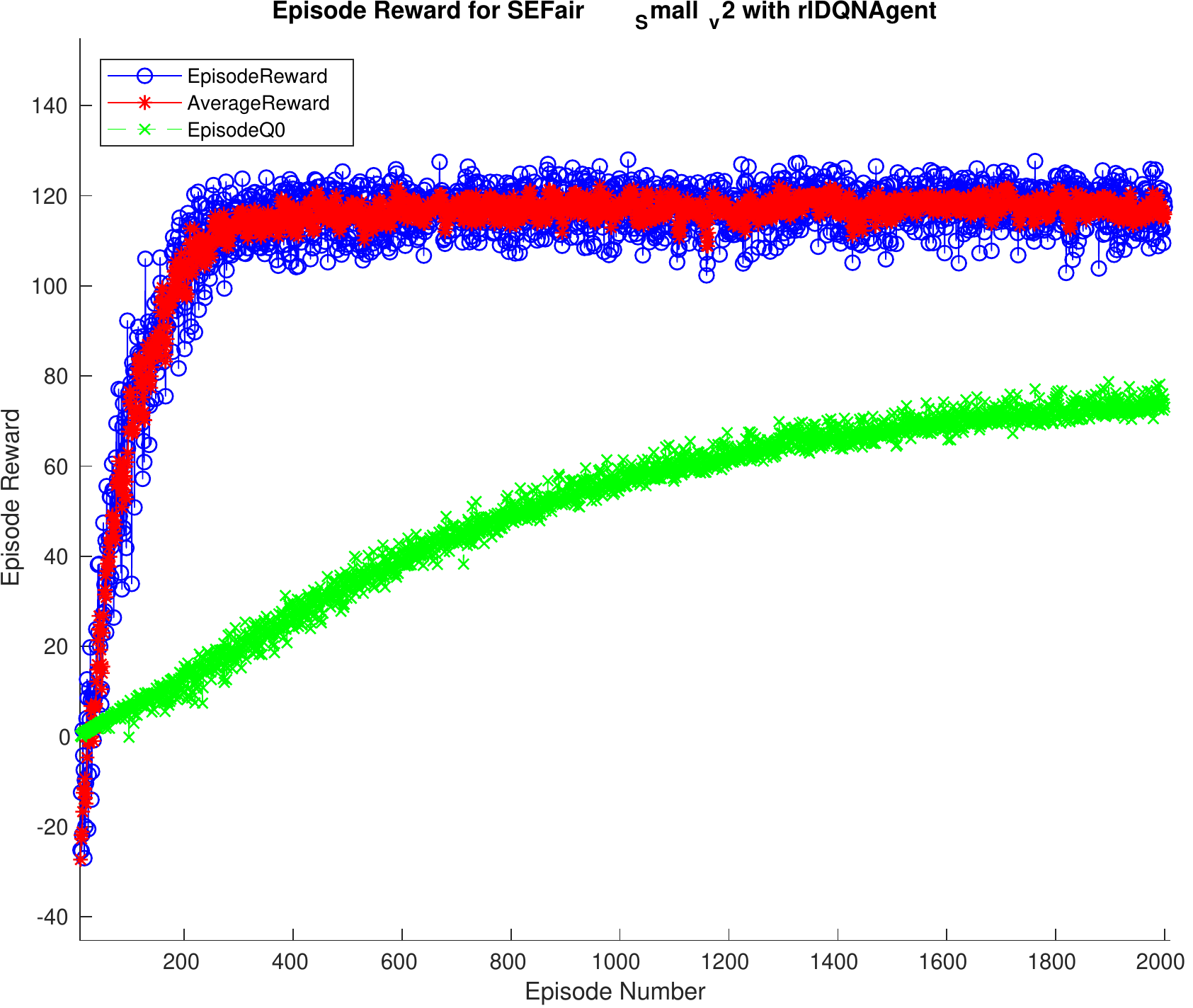}
    \caption{\label{fig:learn_curve} LEASCH learning curve for 2000 episodes. }
  \end{figure}

 	In the training phase, LEASCH is trained in a pool of parallel threads in the GPU. As shown in Figure \ref{fig:learn_curve}, LEASCH was able to converge in less than 300 episodes. The theoretical (long term) reward, the green line in the graph, has also shown a steady increase which indicates a stable learning of LEASCH with each episode. In addition, the average reward (averaged each 5 episodes) has revealed a stable experience by the agent.

  \begin{table}[t]
    \begin{tabular}{p{0.25\columnwidth}|p{0.15\columnwidth} |p{0.4\columnwidth}}
    \hline
     Parameter & Value & Description\\
     \hline 
      $\alpha$ & $1e^{-4}$ & DNN learning rate \\
      Optimizer & Adam \\ 
      Gradient threshold & 1     & \\
      $\epsilon$ & 0.99 & $\epsilon$-greedy parameter\\
      $\min_{\epsilon}$ & 0.01 & Min. allowed $\epsilon$ \\
      $\delta_\epsilon$ & $1e^{-4}$ & $\epsilon$ decaying factor \\
      $|\mathcal{R}|$ & $1e^{6}$ & Experience replay memory size \\ 
      $M$ & 64 &  Mini-batch size \\
      $T$ & 20  & Smoothing frequency \\
      $\beta$ & $1e^{-3}$ & Smoothing threshold \\
      $\ell_{\text{episode}}$ & 150 RBG & Episode length\\
      No. of episodes & 500 & Training episodes \\
     \hline
    \end{tabular}\\
    \caption{\label{tbl:leasch_params} Adopted DRL parameters / hyper-parameters.}
  \end{table}

    \begin{table}[t]
    \begin{tabular}{p{0.3\columnwidth}|p{0.60\columnwidth}}
    \hline
     Parameter & Value\\
     \hline 
      Radio access tech. & 3GPP 5G NR \\
      Test time & 250 frames \\
      Simulation runs & 100 runs with different deployment scenarios \\
      Numerology index $\mu$ & $\{0, 1, 2\}$ \\ 
      Bandwidth & \{5MHz, 10MHz, 20MHz\} \\
      UEs       & 4 \\
      SCS       & $\{$15kHz, 30kHz, 60kHz$\}$\\
      No. of RBs       & $\{25, 24, 24\}$ see \cite{TS_38_101_1} \\
      Scheduling period & 1 RGB \\
      RBG size & 2 RBs according to configuration 1 in \cite{TS_38_214} \\
      Total tested RBGs & $250 \times 100 \times \{130, 240, 480 \}$ RBGs \\
      Channel development & Randomly changes each $\frac{1}{4}$ second \\
      HARQ      & True \\
     \hline
    \end{tabular}\\
    \caption{\label{tbl:simulator_params} Adopted parameters for LEASCH testing on 5G network.}
  \end{table}

\subsubsection{Key performance indicators}
	Throughput, goodput, and fairness are the main key performance indicators (KPIs) used for evaluating the current work. For throughput, the sum of achievable data rate in the cell is reported. For goodput, the delivered data rate is measured at the receiver. For fairness, the popular Jain's fairness index (JFI) is used. 

 \subsection{Ability to solve RRS}
	LEASCH and the baseline algorithms have been tested on different channel bandwidths: 5, 10 and 20 MHz; and different numerology indexes with: 15, 30 and 60 kHz SCS. See Table \ref{tbl:simulator_params}. 

\begin{figure*}[t]
    \centering
    \subfloat[Throughput]{\includegraphics[width=0.5\columnwidth]{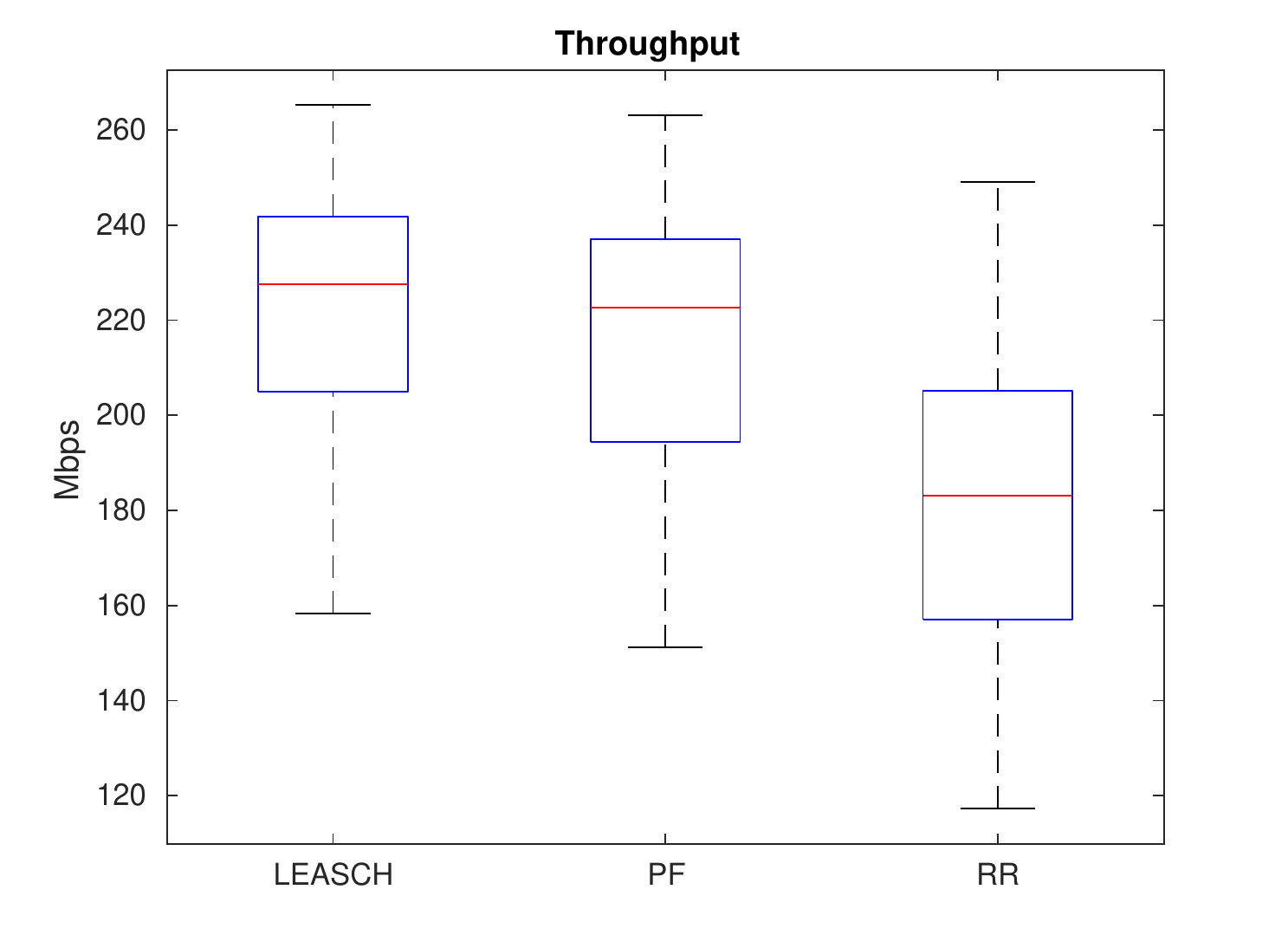}}
    \subfloat[Goodput]{\includegraphics[width=0.5\columnwidth]{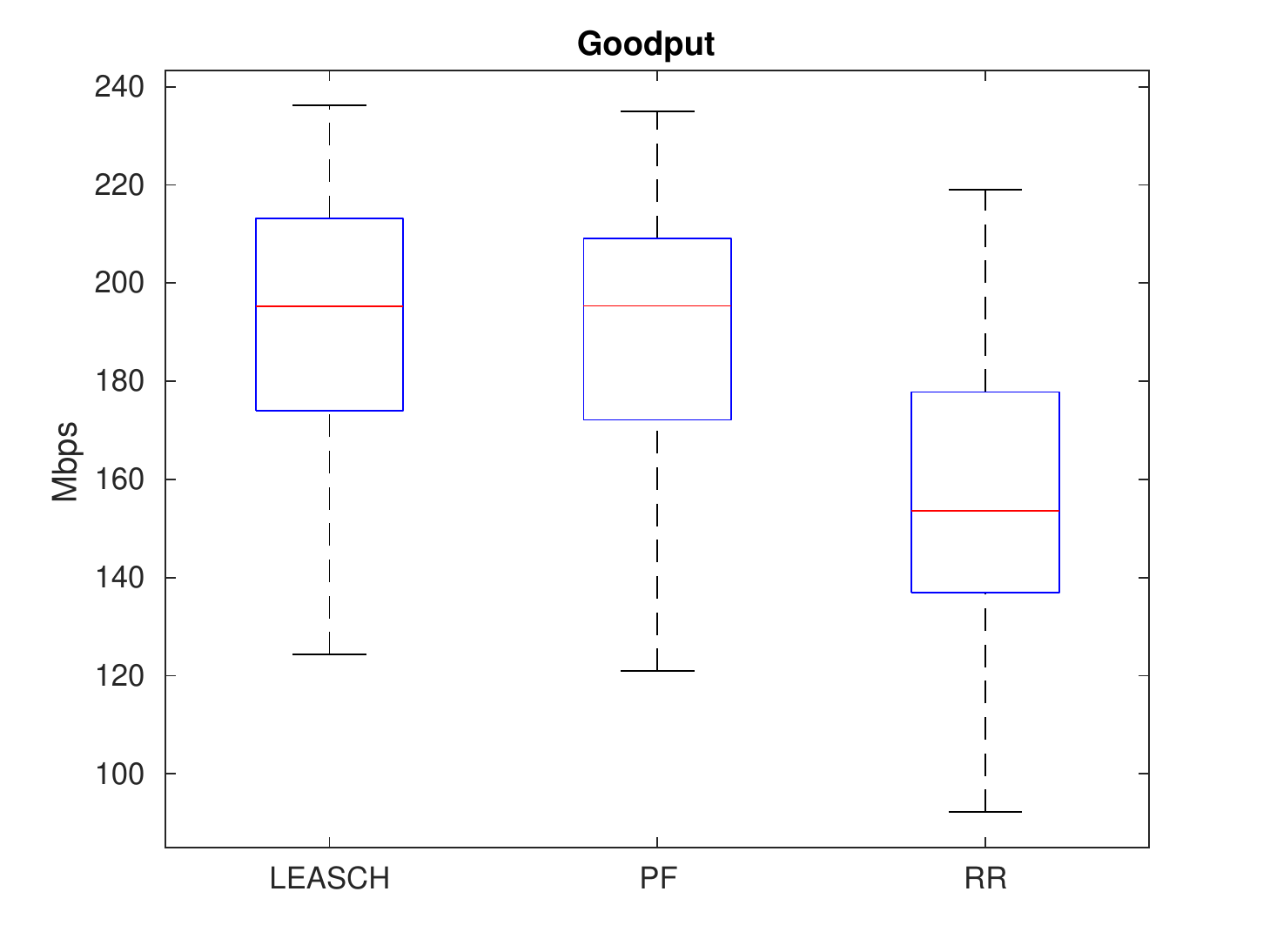}} 
    \subfloat[Fairness]{\includegraphics[width=0.48\columnwidth, trim=1cm 0cm 0cm 0cm, clip]{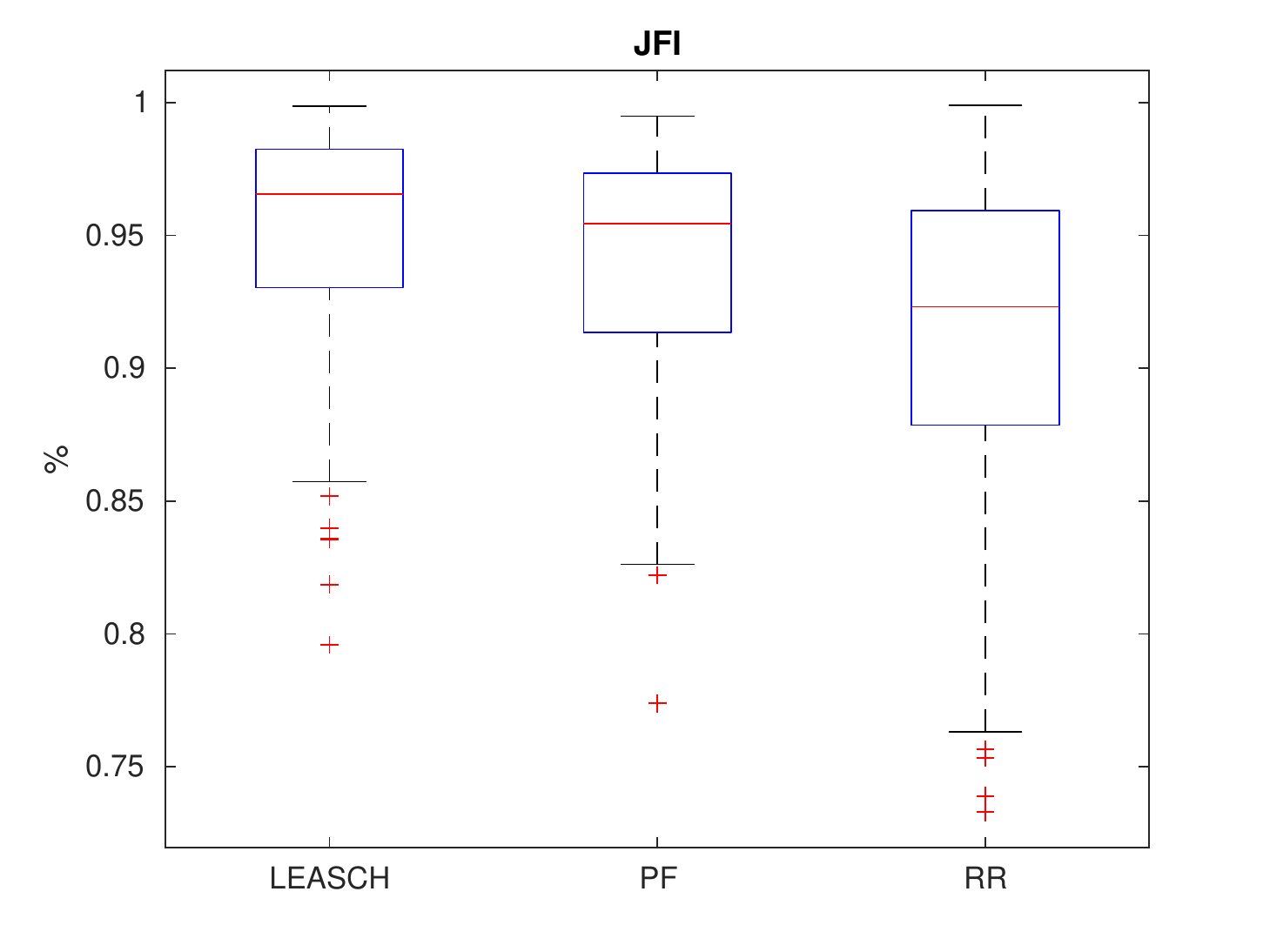}}
    \subfloat[Mean throughput-goodput]{\includegraphics[width=0.5\columnwidth]{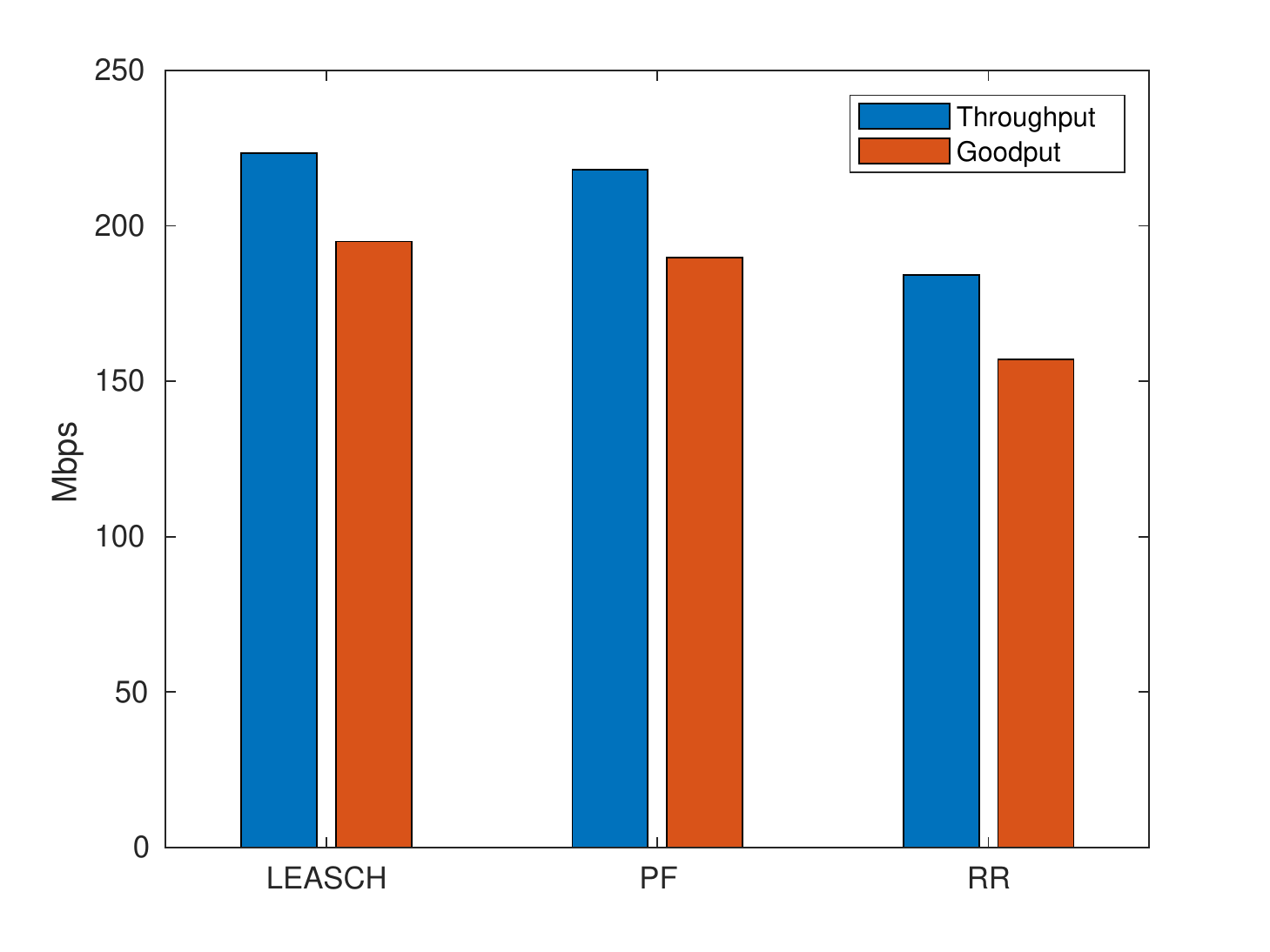}}
    \caption{\label{fig:kpis_5_15khz} KPIs for 250 frames of 15kHz SCS under 5MHz BW for 100 runs.}
  \end{figure*}

	The results of the first group of settings, \ie 5 MHz BW and 15 kHz SCS, are shown in Figure \ref{fig:kpis_5_15khz}. These results clearly demonstrate that LEASCH is better than the baseline in all KPIs. LEASCH has improved the throughput by $\approx$ 2.4\% and 18\% compared to PF, and RR, respectively. In terms of goodput, LEASCH is better by $\approx$ 3\% and 20 \% compared to PF and RR, respectively, which indicates a better stability in LEASCH performance when compared to the baseline. For the JFI, LEASCH is $\approx$ 1\% and 4.3\% better than PF and RR, respectively. 

	For the second set of settings, \ie 10MHz BW and 30kHz SCS, LEASCH has improved the throughput by $\approx$ 3\% and 19\% compared to PF and RR, respectively. In what concerns of goodput, LEASCH is $\approx 3.3\%$ and 21\% better than PF and RR, respectively. Regarding JFI, LEASCH is $\approx$ 2\% and 5\% better than PF and RR, respectively. 

	The third set of settings, \ie 20MHz BW and 60Khz SCS, has also shown similar improvement where LEASCH has improved the throughput by $\approx$ 3\% and 18\% compared to PF and RR, respectively. For goodput, LEASCH outperformed PF and RR by $\approx$ 4\% and 20\%, respectively. Regarding JFI, LEASCH improved the fairness compared to PF and RR by $\approx$ 2\% and 5\%, respectively.

	These results have clearly shown that LEASCH has a competitive and consistent performance compared to the baseline. LEASCH has shown improvement in all measurements, which is not an easy task given that LEASCH has a simple design and has been trained off-simulator. In addition, when choosing a setting with higher theoretical throughput (\eg 10MHz with 30kHz SCS instead of 5MHz with 15kHz SCS), LEASCH was able to scale well and improve the performance even further. One nice property of LEASCH is that it is able to push all the KPIs without compromising any of them. More specifically, LEASCH was able to improve the throughput but at the same time without compromising the goodput. This is why the goodput is enhanced even more than the throughput in all tests compared to the baseline.

  \begin{figure*}
    \centering
    \subfloat[Throughput]{\includegraphics[width=0.50\columnwidth]{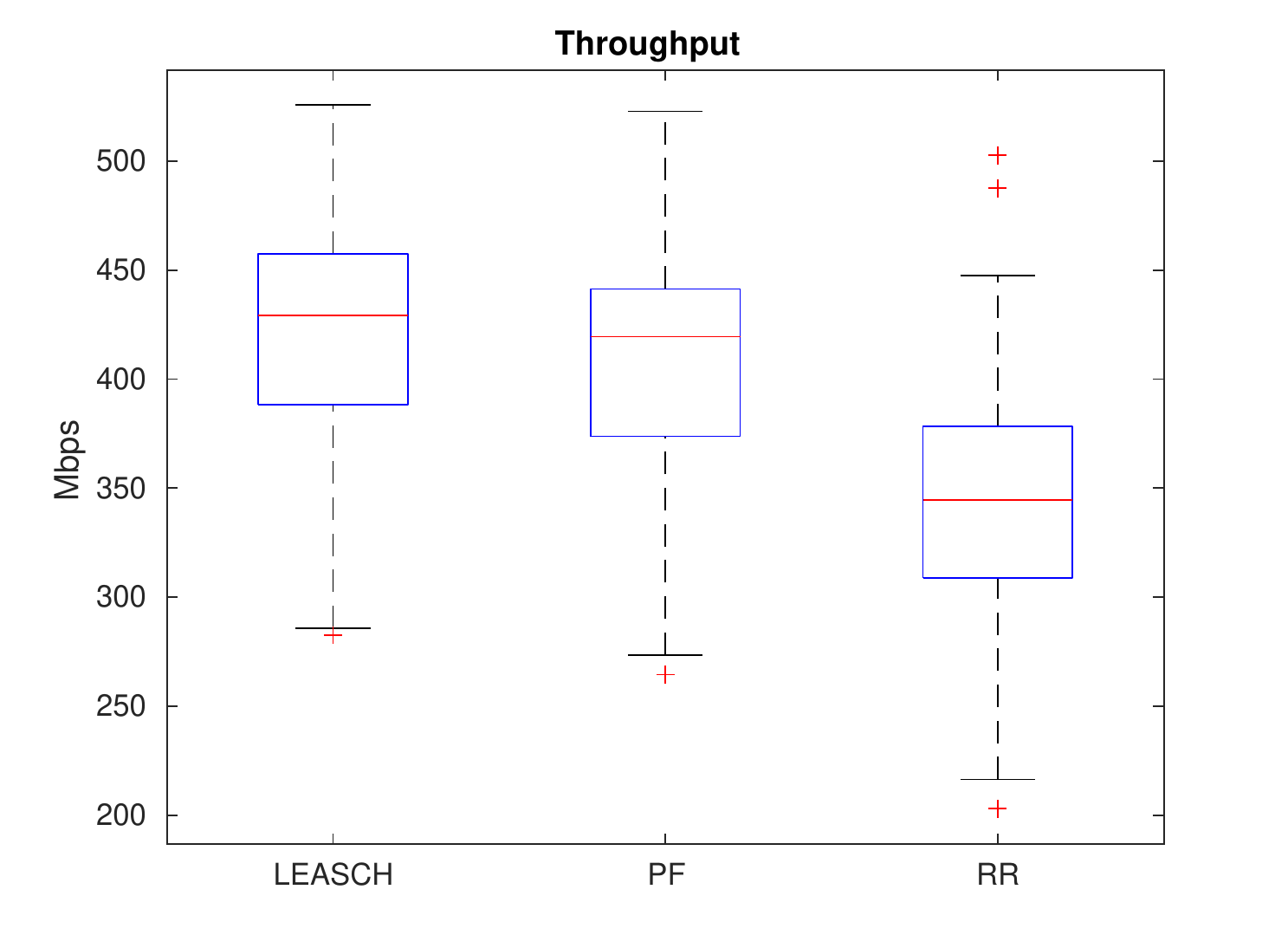}}
    \subfloat[Goodput]{\includegraphics[width=0.50\columnwidth]{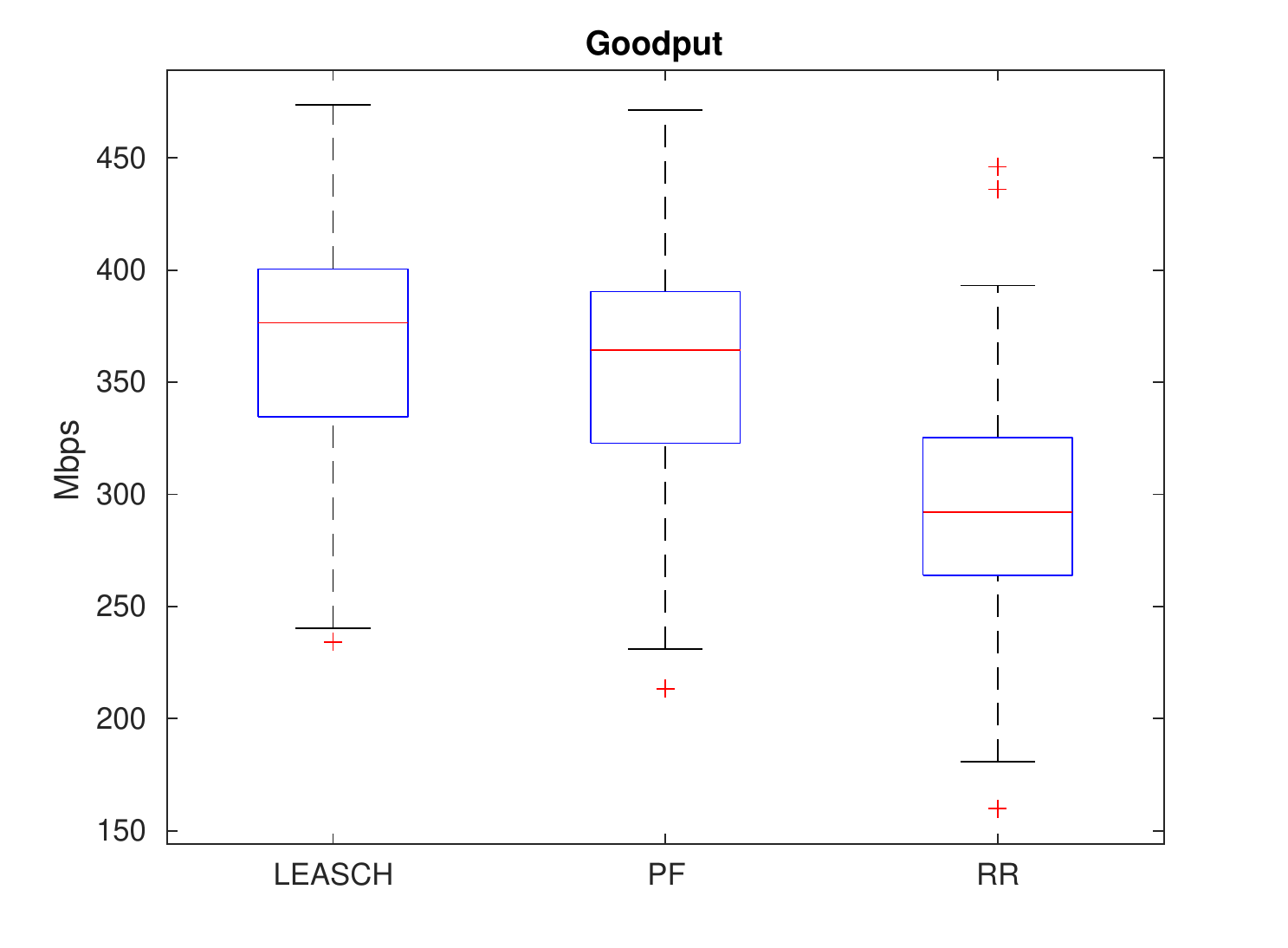}} 
    \subfloat[Fairness]{\includegraphics[width=0.48\columnwidth, trim=1cm 0cm 0cm 0cm, clip]{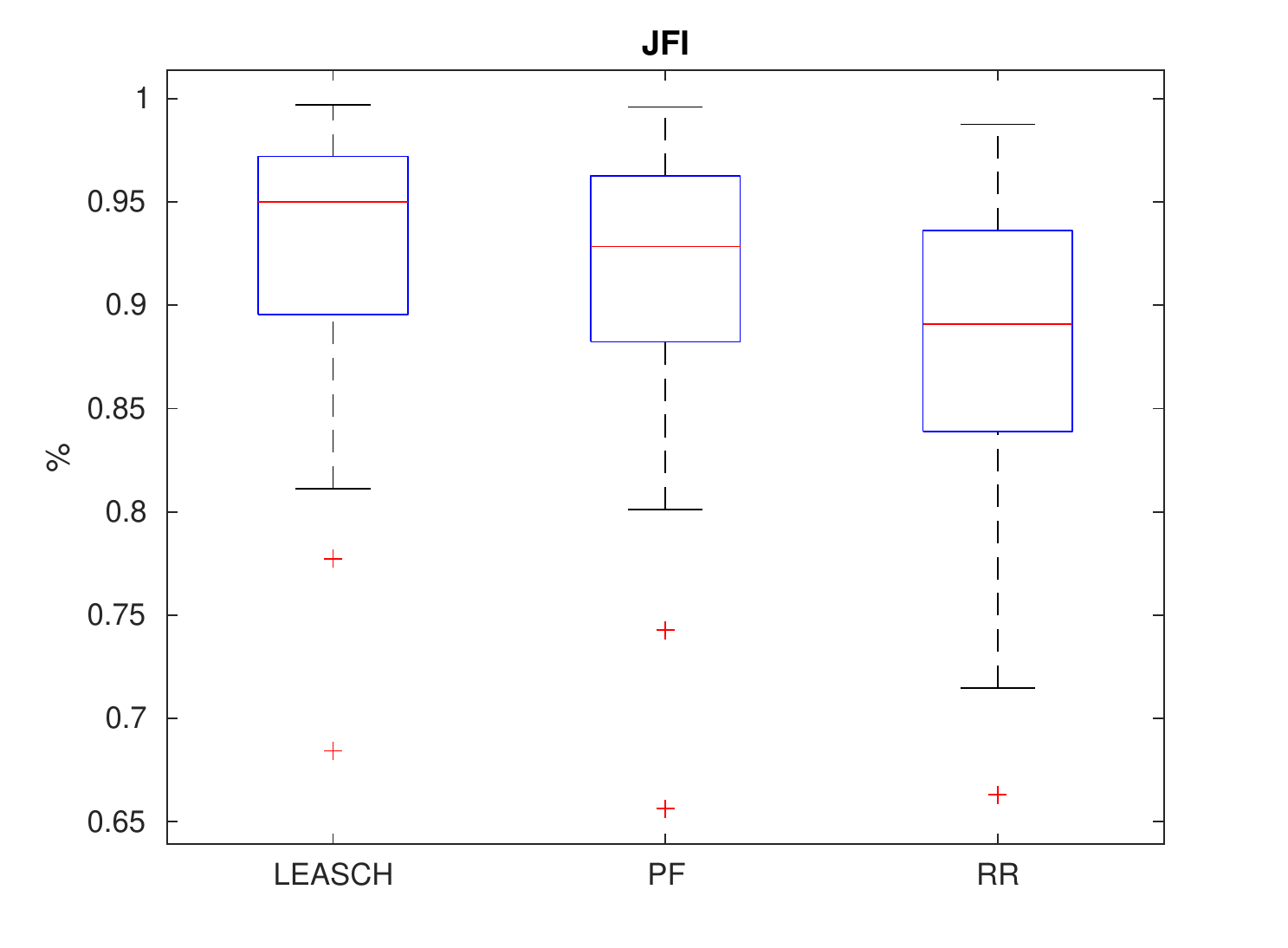}}
    \subfloat[Mean throughput-goodput]{\includegraphics[width=0.50\columnwidth]{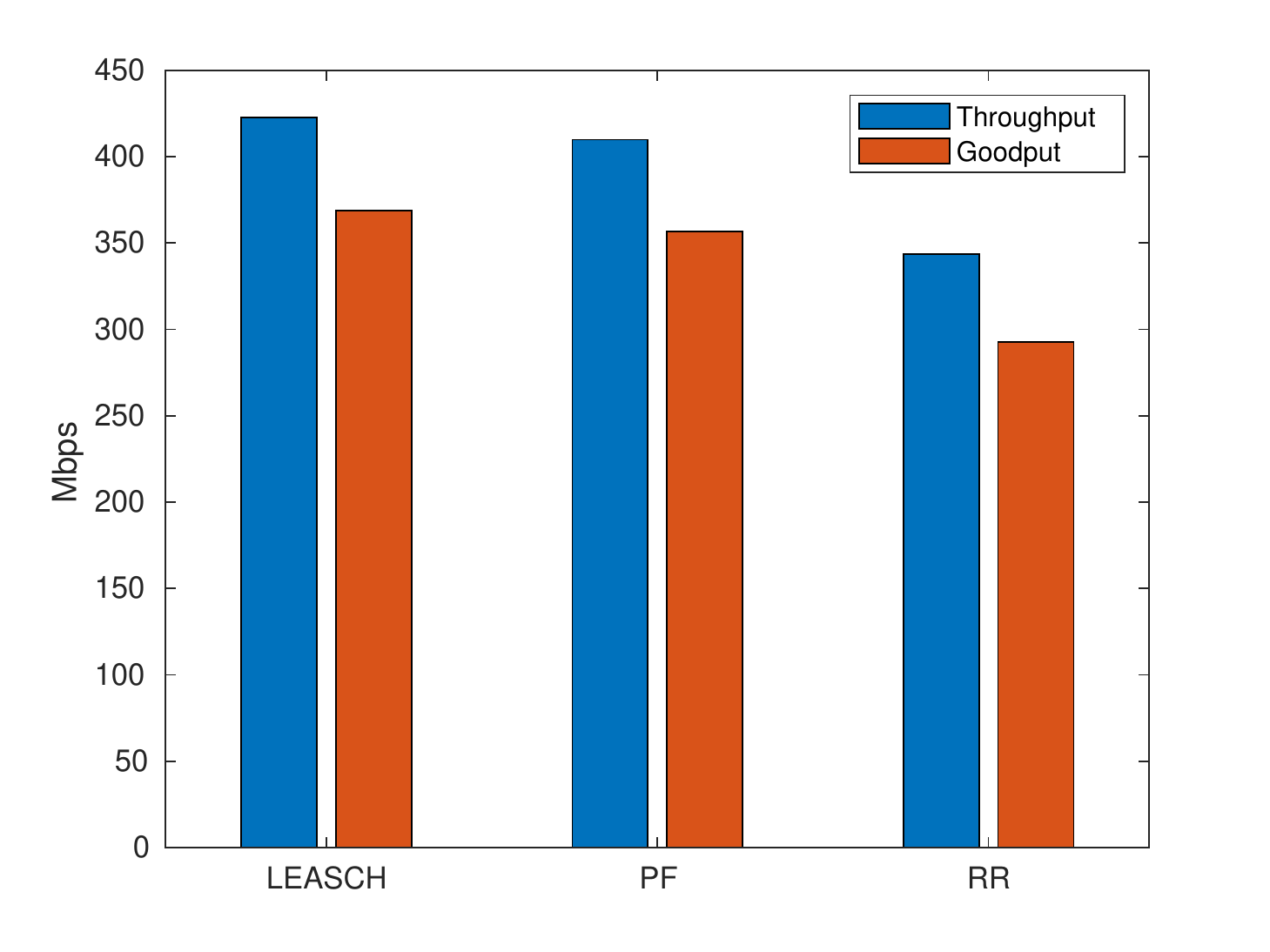}}
    \caption{\label{fig:kpis} KPIs for 250 frames of 30kHz SCS under 10MHz BW for 100 runs.}
  \end{figure*}

  \begin{figure*}
    \centering
    \subfloat[Throughput]{\includegraphics[width=0.50\columnwidth]{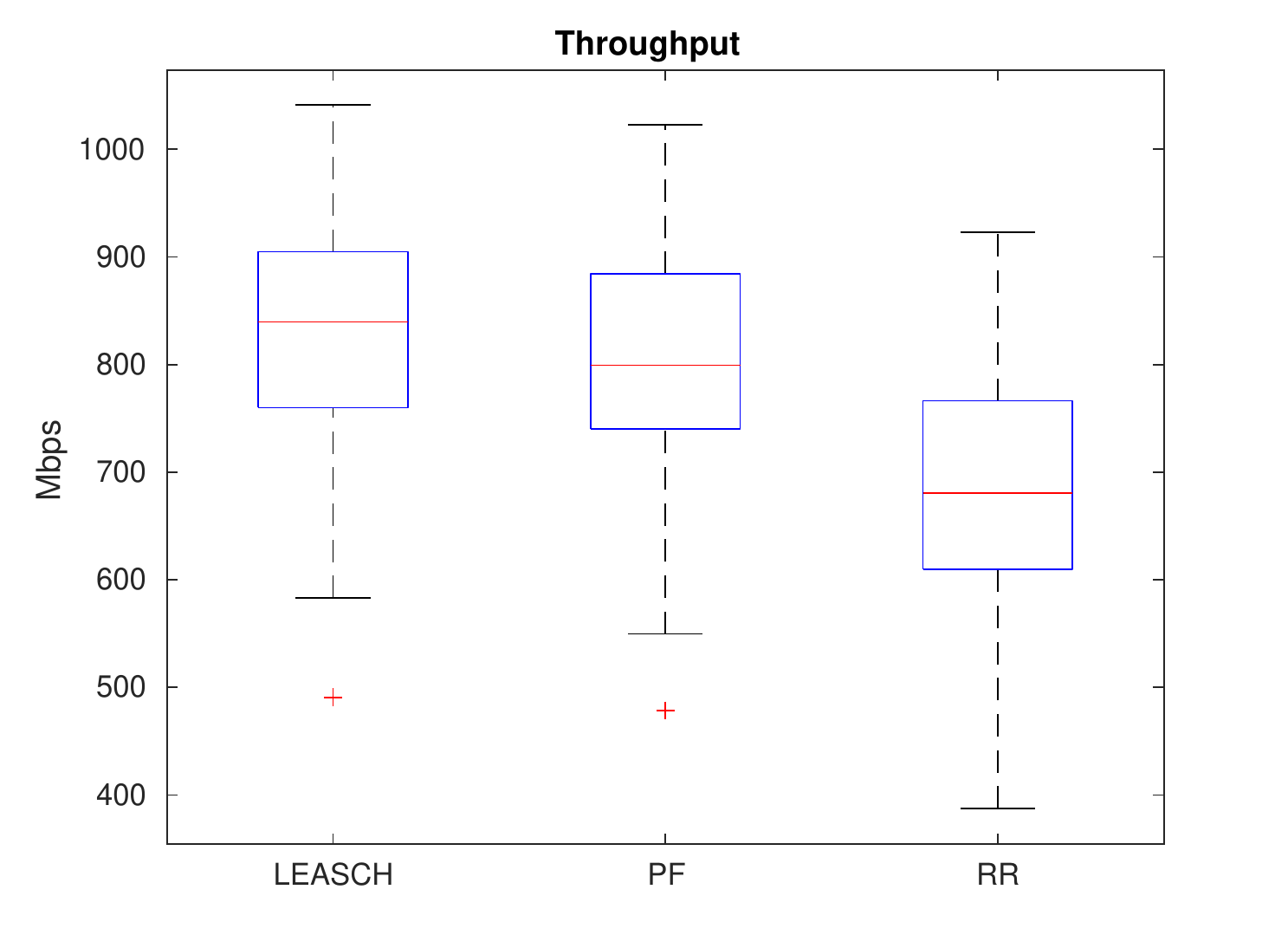}}
    \subfloat[Goodput]{\includegraphics[width=0.50\columnwidth]{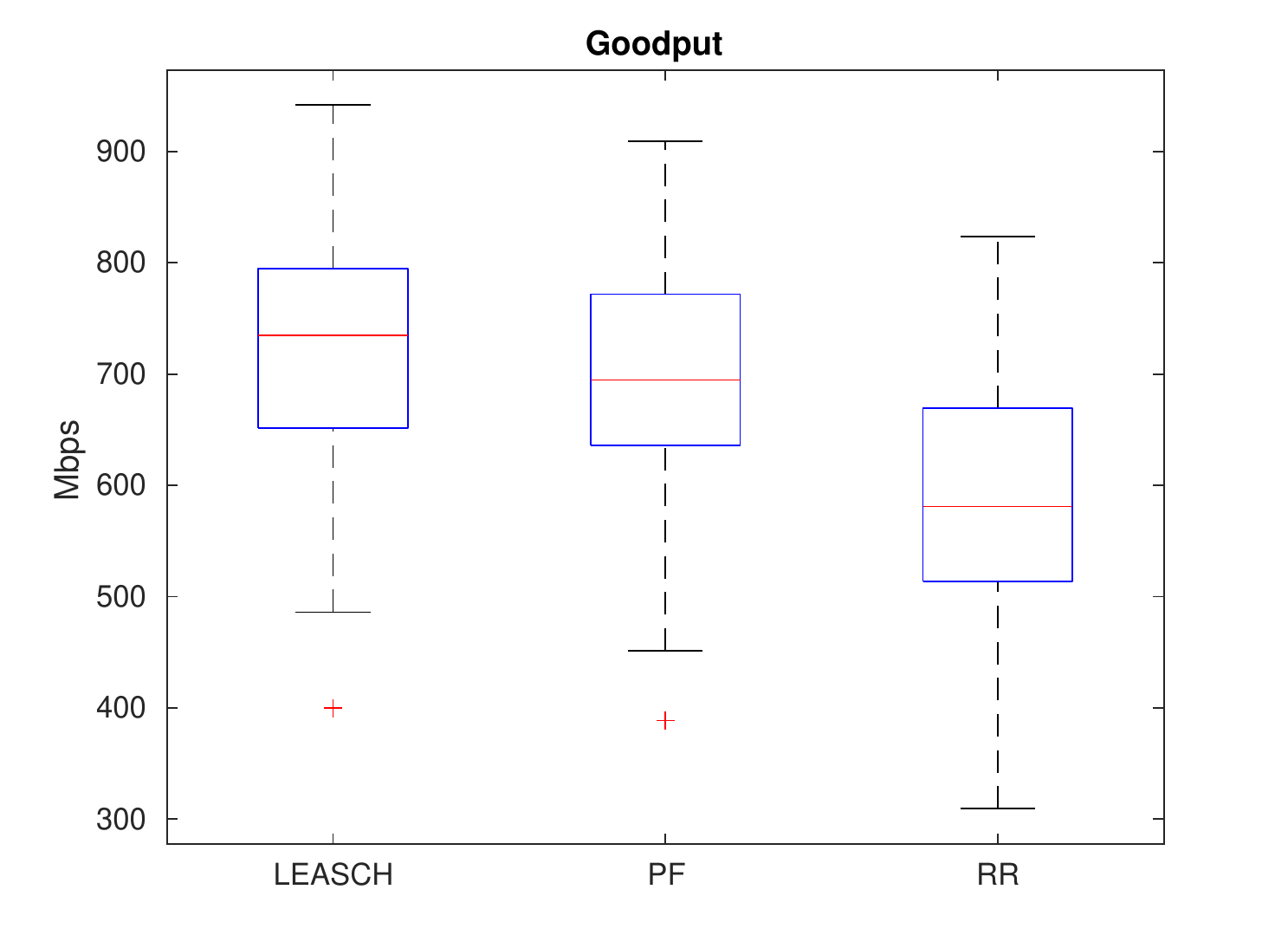}} 
    \subfloat[Fairness]{\includegraphics[width=0.48\columnwidth, trim=1cm 0cm 0cm 0cm, clip]{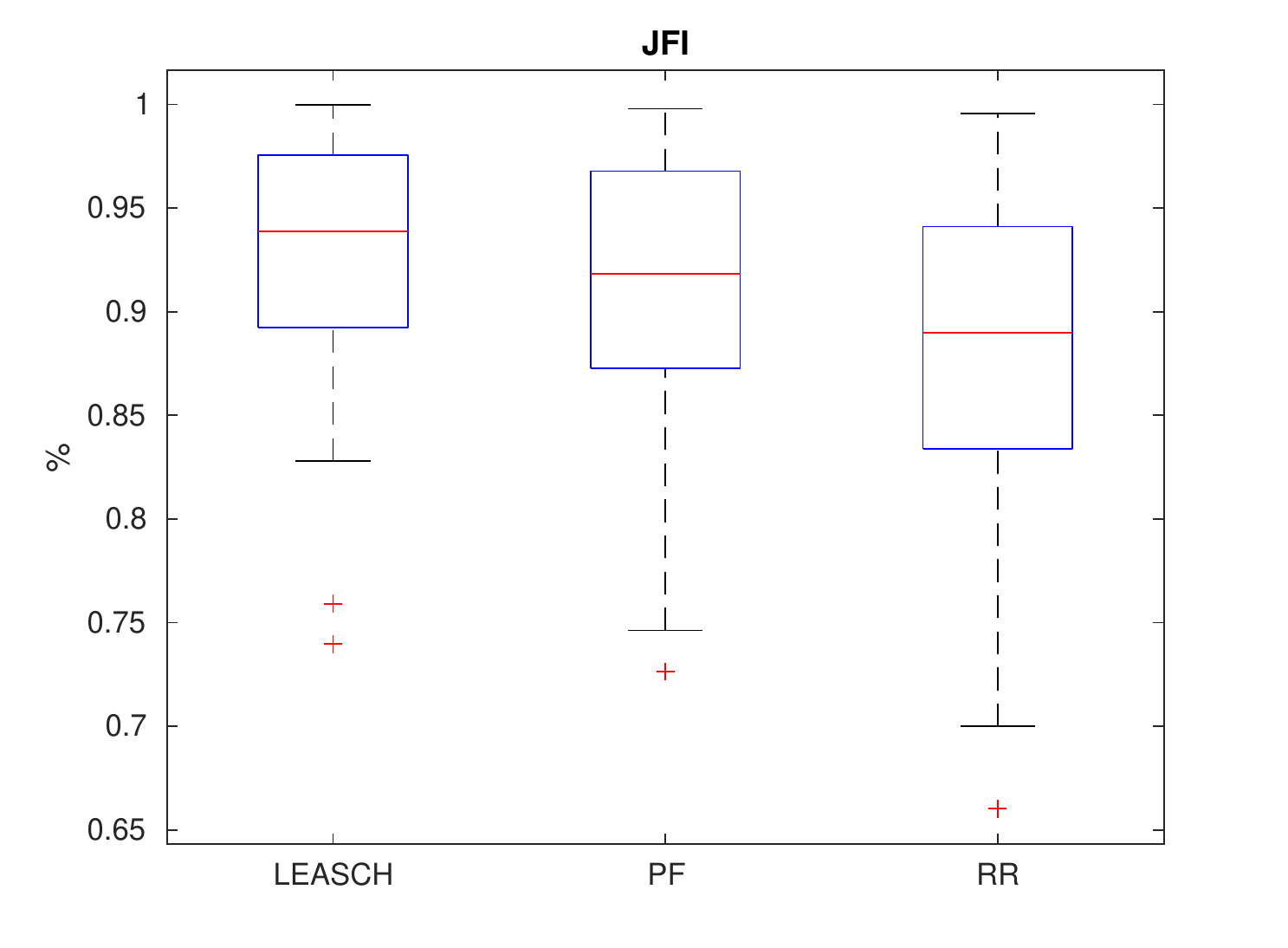}}
    \subfloat[Mean throughput-goodput]{\includegraphics[width=0.50\columnwidth]{figures/mean_th_gd_30khzI.pdf}}
    \caption{\label{fig:kpis} KPIs for 250 frames of 60kHz SCS under 20MHz BW for 100 runs.}
  \end{figure*}

  \subsection{Which policy did LEASCH learn?}
    This section tries to analyze and figure out which policy did LEASCH learn. This task is not trivial, not only for LEASCH but for almost every DRL agent. Here it is more difficult not only because of the stochastic nature of LEASCH, but also due to the complexity of the RRS problem. Therefore, the visual inspection approach of LEASCH behavior will be followed.

\begin{figure*}
    	\subfloat[LEASCH]{
    					 \includegraphics[width=0.99\columnwidth]{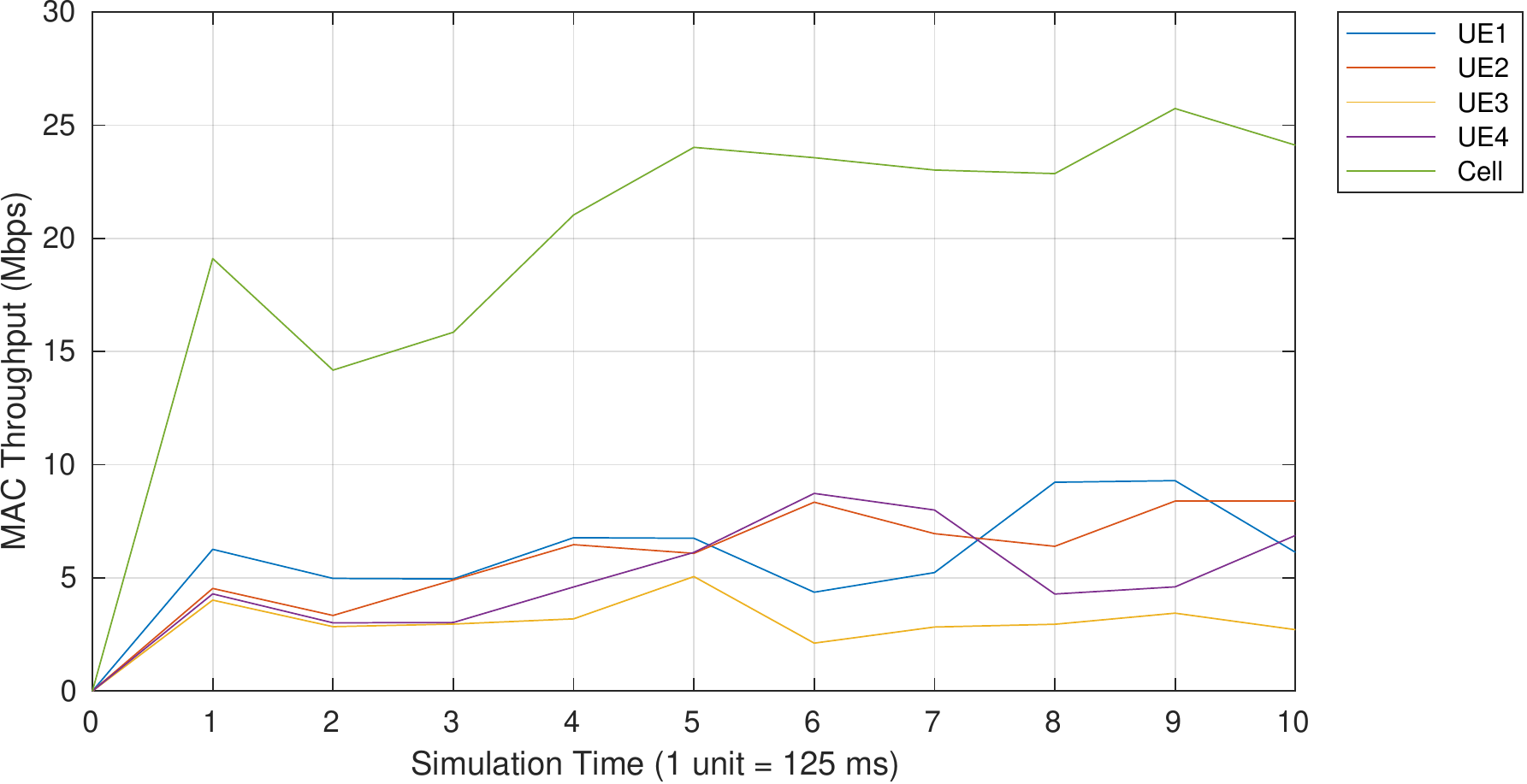}
    					 \includegraphics[width=0.99\columnwidth]{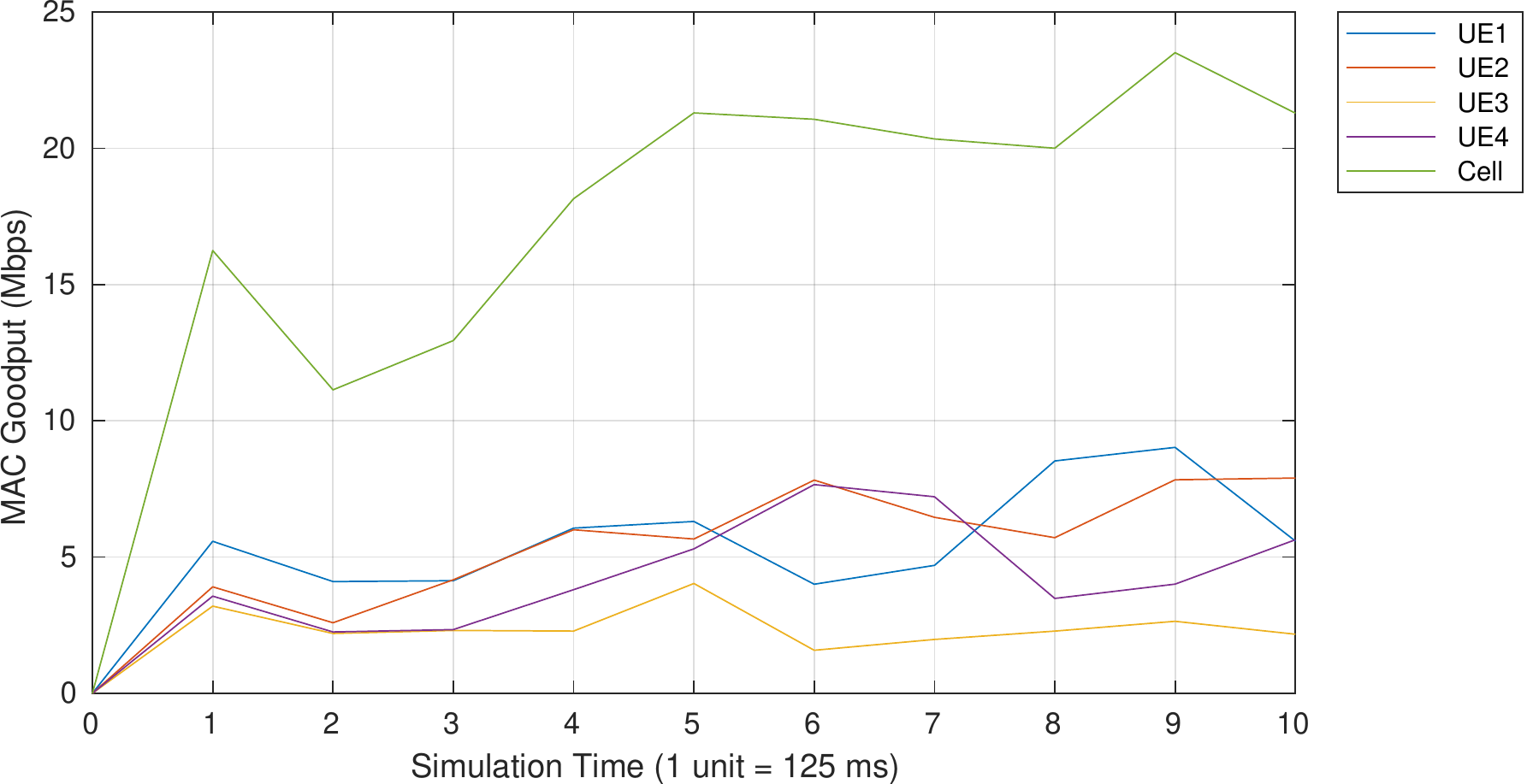}
    					 } \\
    	\subfloat[PF]{
    					 \includegraphics[width=0.99\columnwidth]{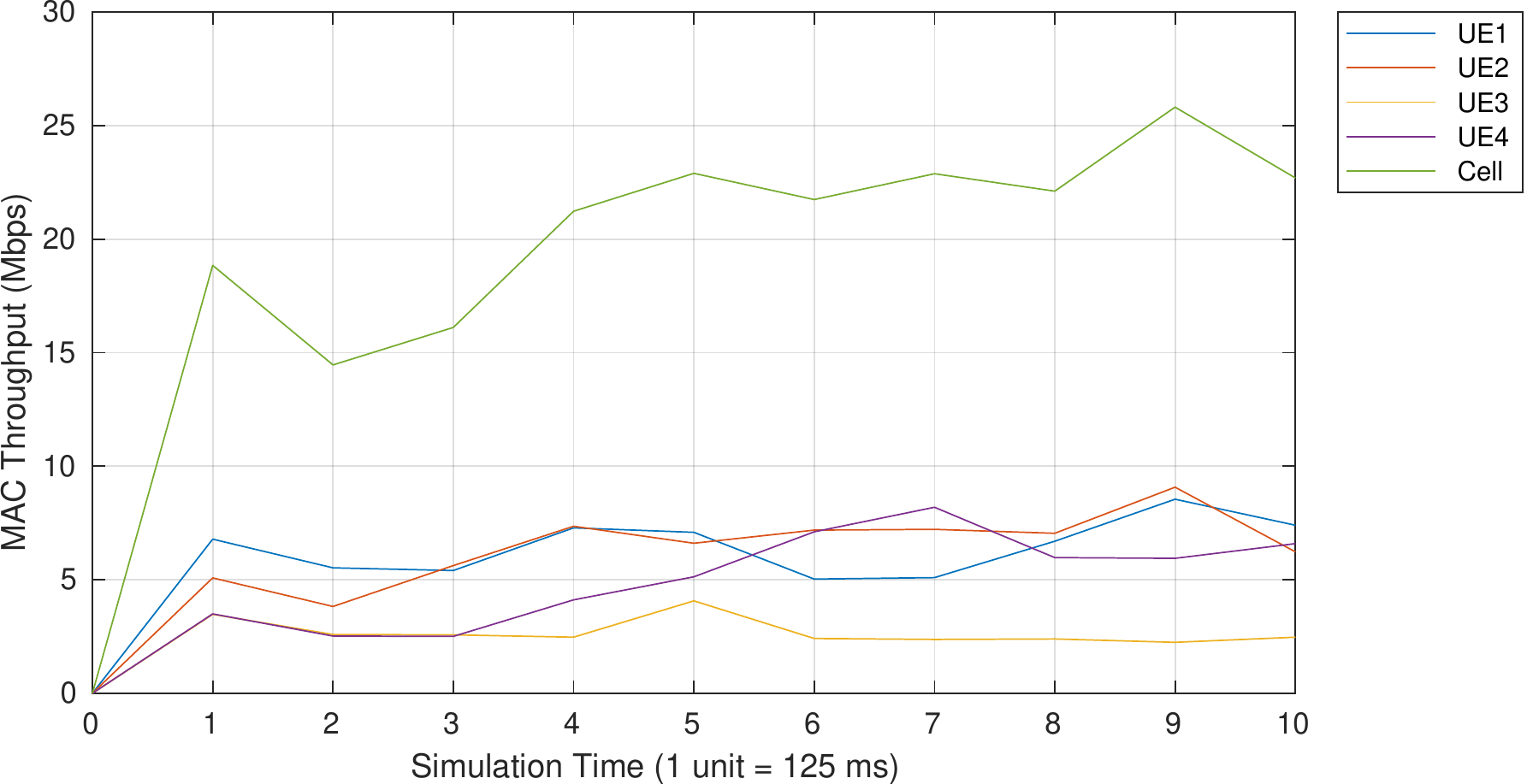}
    					 \includegraphics[width=0.99\columnwidth]{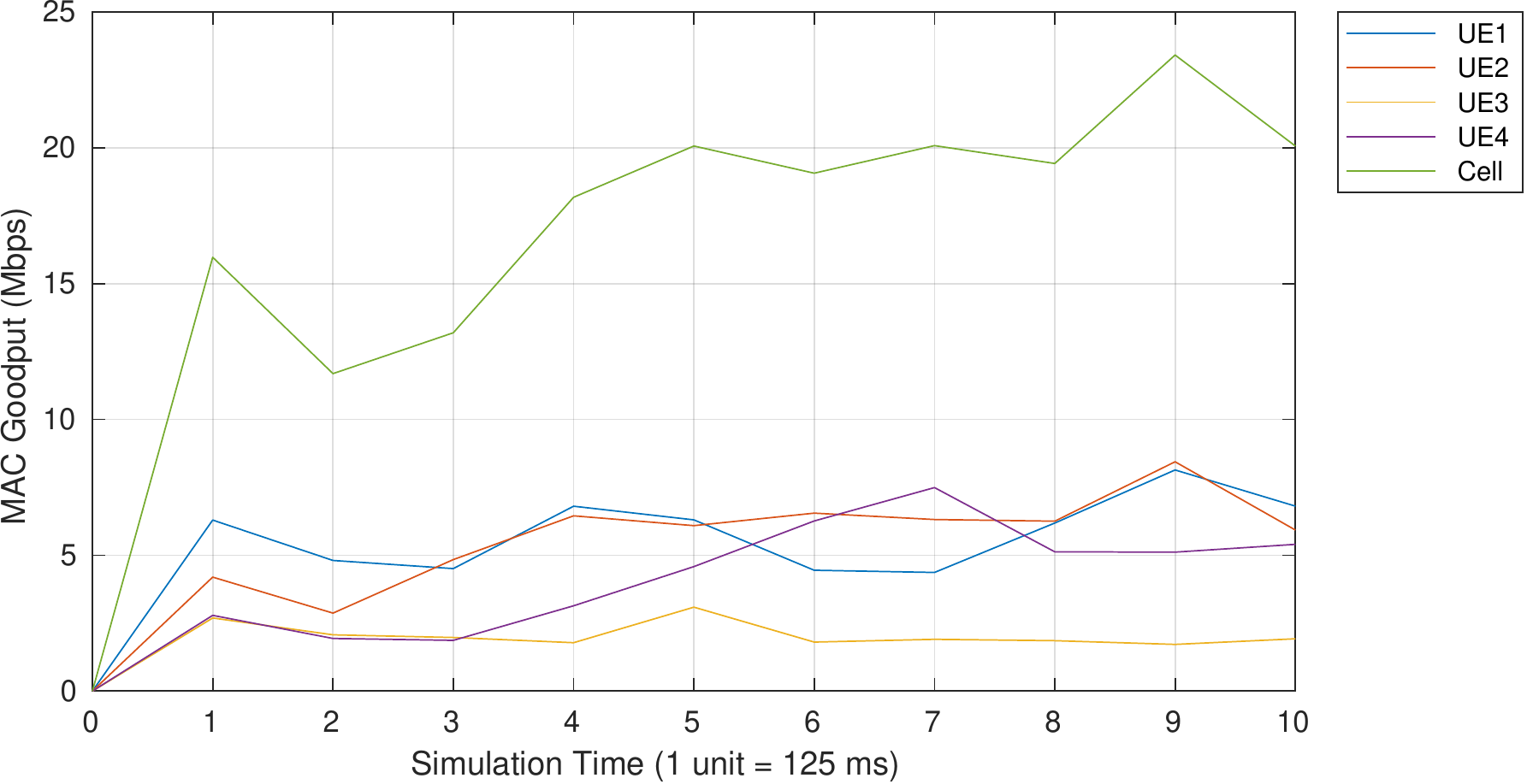}
    				} \\

    	\subfloat[RR]{
    					 \includegraphics[width=0.99\columnwidth]{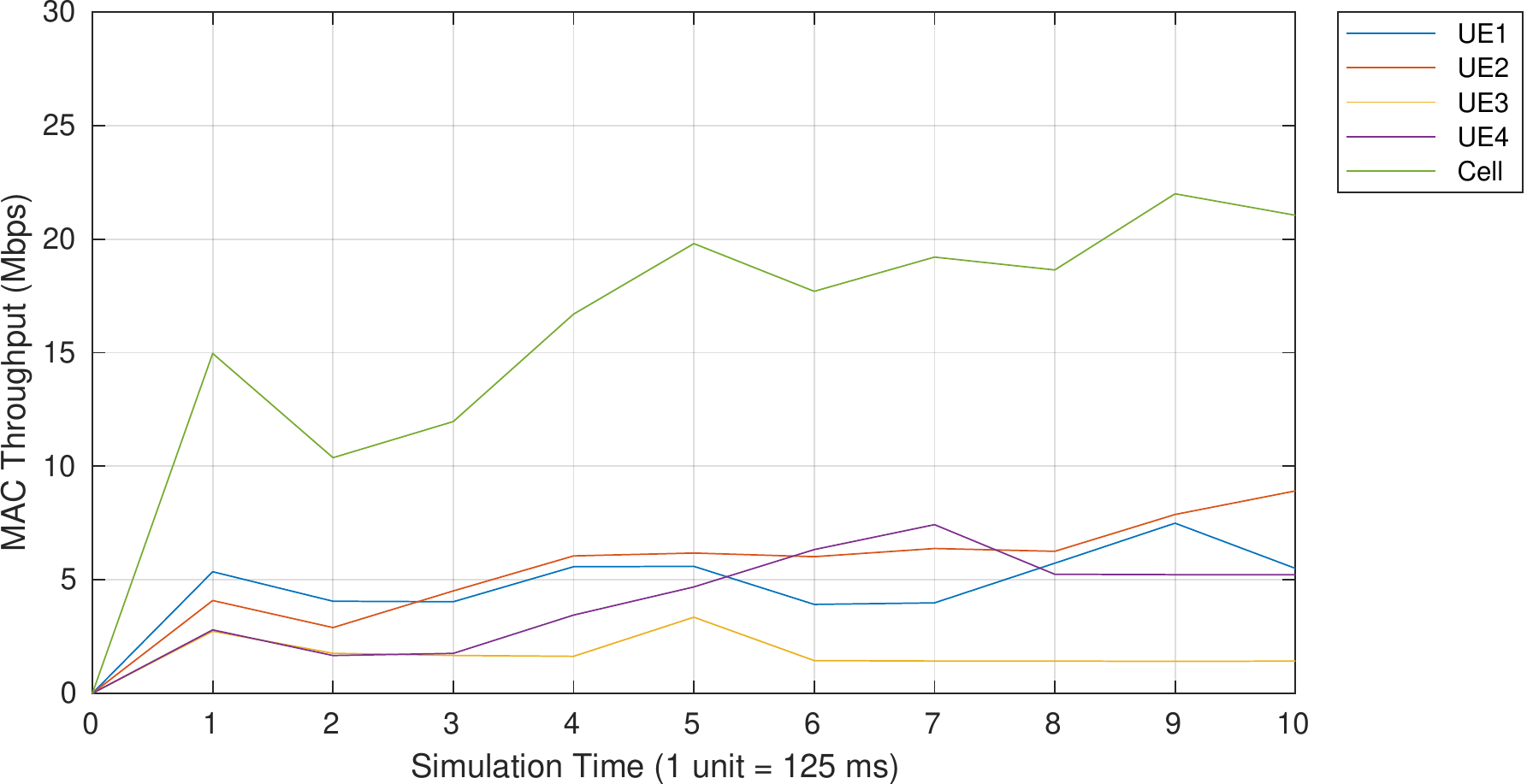}
    					 \includegraphics[width=0.99\columnwidth]{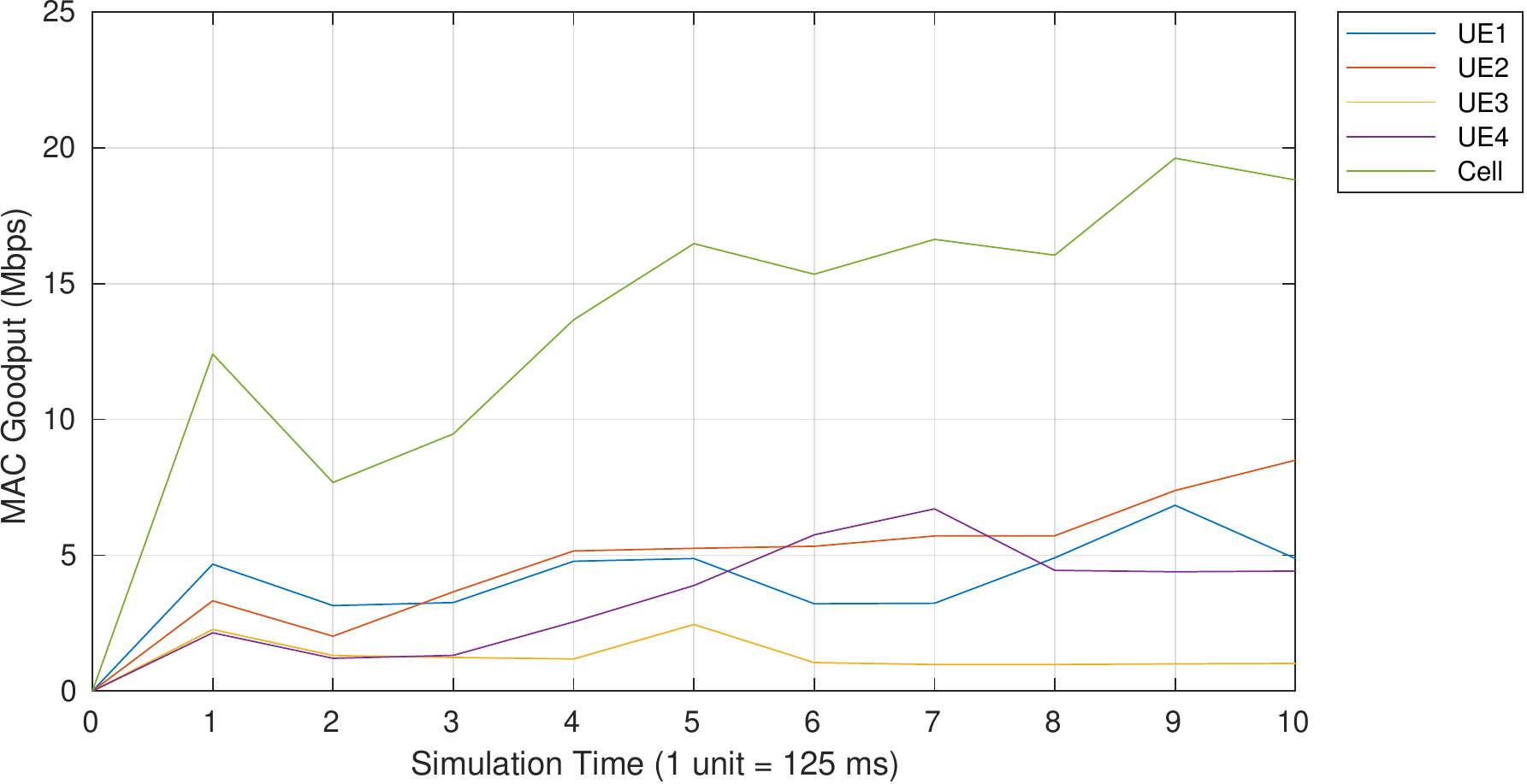}
    				}
    	\caption{\label{fig:30khz_curves} A random testing run for the 10MHz BW and 30kHz SCS setting. Left column: throughput; right column: goodput.}
    \end{figure*}

    To that end, a testing run is sampled for a set of settings and the throughput and goodput curves are quantized into 10 time units (see Figure \ref{fig:30khz_curves}). These curves are then visually inspected with regard to those of PF and RR. By comparing these curves, both for each UE and for the cell, it is possible to construct an idea about which policy LEASCH has learned. In this figure, the second set of settings with 10MHz BW and 30kHz SCS is chosen. Since 30kHz SCS is used, the simulation time is only 1250 ms. For 15kHz this would be 2500 ms. This is due to the reduction in symbol duration as the numerology index increases. 

    According to the simulation settings in Table \ref{tbl:simulator_params}, the channel changes (and consequentially new  CQI feedbacks are signaled) each 125 ms, \ie each 1 time unit in Figure \ref{fig:30khz_curves}.  In this Figure, first it is interesting to see that the trends of the cell curves are similar in all approaches. However, at each time unit each method makes different scheduling decisions. Second, LEASCH outperforms PF and RR since it reaches higher throughput-goodput, especially from the period 5 to 8 time units where major changes have occurred in the channel. Before time unit 5 (\ie from 1 to 4) LEASCH performed almost identical to PF in terms of throughput-goodput but, at the same time, the UEs' curves are more compact in LEASCH which indicates a better fairness. After the period 5 to 8 time units, LEASCH continued to maintain high throughput without sacrificing UEs that have bad CQIs (\eg compare the curve of UE3 in all approaches). Although our discussion here lacks analytical bases, due to the complexity of the problem, it is clear that LEASCH has nicely realized the intuitions we have designed it for. LEASCH tries to improve all KPIs without sacrificing UEs with bad CQIs, by wisely distributing the spectrum among all UEs.

  \subsection{Learning performance}
  	Here the learning performance of LEASCH is analyzed. The main objective is to asses its design quality given that it has to learn two different goals: avoid scheduling inactive UEs, and joint optimize throughput and fairness. Using only theoretical foundation of DRL, it is not easy to see how LEASCH learned these different (and perhaps contradictorily) goals. The reason is that, the learned weights of the $Q$ networks can not easily be interpreted to assess the learning performance and the quality of LEASCH's design. Therefore, a reward analysis is performed by separating both goals outcomes.

\begin{figure}
  	\centering
  	\includegraphics[width=0.82\columnwidth, trim=1.2cm 0cm 0.5cm 0.5cm, clip]{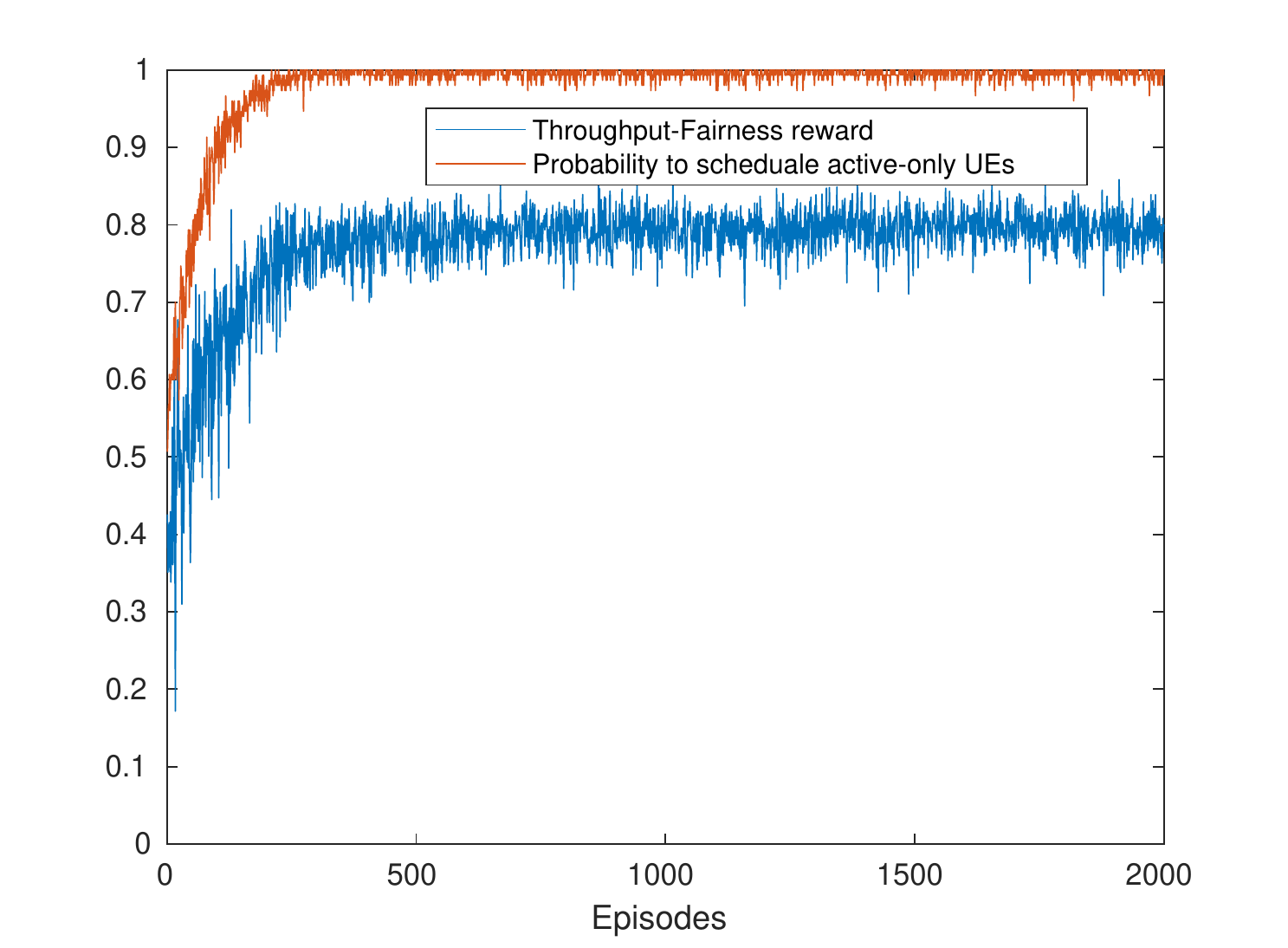}
  	\caption{\label{fig:prob} Learning different objectives by LEASCH.}
  \end{figure}

  	To that end, the learning curve in Figure \ref{fig:learn_curve} is decomposed into two curves as shown in Figure \ref{fig:prob}. In addition, instead of calculating the average total reward of the episode (as in Figure \ref{fig:learn_curve}), the average reward of each episode is used. This allows us to study how LEASCH learns both parts of expression \eqref{equ:reward} separately. From this figure, the red curve represents the probability of scheduling active-only UEs while the blue curve is the throughput-fairness reward, \ie $\left\{\hat{d}_u \times \frac{\min\limits_u f_u}{\max\limits_u f_u}\right\}$ in \eqref{equ:reward}. These two curves show that LEASCH was able to jointly learn these two objectives and, around episode 300, LEASCH was able to converge for both objectives which clearly indicates the effectiveness of LEASCH's design. In addition, this also shows the suitability of DRL to handle the scheduling problem, which is usually a multi-objective problem. 

\section{Conclusions} \label{sec:conclusions}
	It has been presented LEASCH, a deep reinforcement learning agent able to solve the radio resource scheduling problem in 5G. LEASCH is a breed of DDQN critic-only agents that learns discrete actions from a sequence of states. It does so by adapting its neural networks, known as DQNs, weights according to the reward signal it receives from the environment. What makes LEASCH different from conventional schedulers is that, it is able to learn the scheduling task from scratch with zero knowledge about the RRS at all. LEASCH is different from the extremely scarce and new AI-schedulers by many things. First LEASCH is trained off-simulator to break any dependency between learning and deployment phases. Making LEASCH a generic tool in any networking AI-ecosystem. Second, LEASCH has novel design not addressed in earlier approaches. Finally, LEASCH was designed as numerology-agnostic which makes it suitable for 5G deployments. 

	Concerning LEASCH performance, it has been compared to the well-established approaches PF and RR. Despite LEASCH's simple design it has shown clear improvement and stability in throughput, goodput, and fairness KPIs. Further analysis has also shown that, LEASCH is able to learn not only to enhance the classical throughput-fairness tradeoff, but to learn not to schedule inactive users. It was able to learn both objectives at the same time as the learning curves depicted. Another interesting property of LEASCH is that, it avoids to penalize users with bad CQIs and tries to keep all KPIs high at the same time. Such property can be improved in the future. In addition, more interesting properties, that can not be easily obtained by conventional approaches, can be learned by LEASCH.

	As a future work, a more advanced version of LEASCH, LEASCH version 2, will be developed to serve larger set of users. LEASCH, as any other DQN agents, is currently not suitable for large scale networks, since DQN agents are known to compromise the performance as the action space increases. Therefore, in the future work, LEASCH version 2 will be developed and deployed under larger 5G network with a mixture of numerology and different type of services.

  \section{Acknowledgment}
  This work is supported by the European Regional Development Fund (FEDER), through the Competitiveness and Internationalization Operational Programme (COMPETE 2020), Regional Operational Program of the Algarve (2020), Funda\c{c}\~{a}o para a ci\^{e}ncia e Tecnologia; i-Five: Extens\~{a}o do acesso de espectro din\^{a}mico para r\'{a}dio 5G, POCI-01-0145-FEDER-030500. This work is also supported by Funda\c{c}\~{a}o para a ci\^{e}ncia e Tecnologia, Portugal, within CEOT (Center for Electronic, Optoelectronic and Telecommunications) and UID/MULTI/00631/2020 project.






\bibliographystyle{plain} 
\bibliography{../../bibtexFDB} 
\end{document}